%% file: ms.tex
\documentclass[]{emulateapj}
\usepackage{graphicx}
\usepackage{amsmath}

\newcommand{\ba}{\begin{array}}
\newcommand{\ea}{\end{array}}
\newcommand{\ben}{\begin{enumerate}}
\newcommand{\een}{\end{enumerate}}
\newcommand{\bei}{\begin{itemize}}
\newcommand{\eei}{\end{itemize}}

\newcommand{\f}[2]{\frac{#1}{#2}}

\newcommand{\rhob}{\rho_{\text{break}}}
\newcommand{\pb}{p_{\text{break}}}
\newcommand{\vrhob}{v_{\rho\text{break}}}
\newcommand{\vpb}{v_{p\text{break}}}
\newcommand{\TbHe}{1.07\, \eV }

\newcommand{\dmbreak}{0.1}
\newcommand{\dmsb}{\delta_{m,\rm BO}}
\newcommand{\leff}{\lambda_{\text{eff}}}
\input{Definitions.tex}

\begin{document}
\title{The early UV/Optical emission from core-collapse supernovae}
\author{Itay Rabinak and Eli Waxman}
\affiliation{Department of Particle Physics and Astrophysics, The Weizmann Institute of Science, Rehovot 76100, Israel}

\email{itay.rabinak@weizmann.ac.il}
\date{\today}

\begin{abstract}
We derive a simple approximate model describing the early, hours to days, UV/optical supernova emission, which is produced by the expansion of the outer $\lesssim10^{-2}M_\odot$ part of the shock-heated envelope, and precedes optical emission driven by radioactive decay. Our model includes an approximate description of the time dependence of the opacity (due mainly to recombination), and of the deviation of the emitted spectrum from a black body spectrum. We show that the characteristics of the early UV/O emission constrain the radius of the progenitor star, $R_*$, its envelope composition, and the ratio of the ejecta energy to its mass, $E/M$. For He envelopes, neglecting the effect of recombination may lead to an over estimate of  $R_*$ by more than an order of magnitude. We also show that the relative extinction at different wavelengths ($A_\lambda-A_V$) may be inferred from the light-curves at these wave-lengths, removing the uncertainty in the estimate of $R_*$ due to reddening (but not the uncertainty in $E/M$ due to uncertainty in absolute extinction). The early UV/O observations of the type Ib SN2008D and of the type IIp SNLS-04D2dc are consistent with our model predictions. For SN2008D we find $R_*\approx10^{11}$~cm, and an indication that the He envelope contains a significant C/O fraction.


\end{abstract}

\maketitle

\section{Introduction}
\label{sec:introduction}

During the past four years, the wide field X-ray detectors on board the Swift satellite enabled us, for the first time, to "catch" supernova (SN) explosions very close to the onset of the explosion. In two cases, SN2006aj and SN2008D, the usual  optical SN light curve was observed to be preceded by a luminous X-ray outburst, which has triggered the X-ray detectors,  followed by a longer, $\sim1$~day, duration UV/O emission \citep{EliNatur,Alicia08D}. Analysis of the later optical SN emission revealed that both were of type Ib/c, probably produced by compact Wolf-Rayet (hereafter WR) progenitor stars, which lost most of their Hydrogen envelope
\citep{Pian06Nat,Mazzali06Nat,Modjaz07,Maeda07Apj,Mazzali08,Malesani09}.
Following the detection of the early UV/O emission from these SNe, a search was conducted for UV/O emission from SNe that fall within the deep imaging survey of the Galaxy Evolution Explorer (GALEX) space telescope, leading to the detection of early ($\sim1$day) rising UV/O emission for two type II-p SNe \citep{Gezari08,Schawinski08}, for which the progenitors are likely red super giants (hereafter RSG) with large Hydrogen envelopes.

X-ray outbursts followed by early UV emission have long been expected to mark the onset of SN explosions.
The SN shock wave, which travels through, and ejects, the stellar envelope, becomes radiation-mediated when propagating through the envelope \citep[for review see, e.g.,][]{WW86}. As the shock propagates outward, the (Thomson) optical depth of the plasma lying ahead of it decreases. When this optical depth becomes comparable to the shock transition optical depth, $\tau_s \simeq c/v_s$ ($v_s$ is the shock velocity), the radiation escapes ahead of the shock. This leads to an expected "shock breakout" X-ray flash \citep{Colgate74,Falk78,Klein78} lasting for 10's to 100's of seconds.
Following breakout, the stellar envelope expands and cools (nearly adiabatically).
As the photosphere penetrates into the outer shells of the envelope, the (adiabatically cooled) radiation stored within the envelope escapes, leading to an expected early UV/O emission \citep{Falk78}. In this paper we focus on the  early , $\sim1$~day, part of this UV/O emission (although it may dominate the total emission for much longer, e.g. in type II-P SNe).

The interpretation of SN associated X-ray outbursts as due to "shock breakout" is not generally accepted, due mainly to the fact that while a $\sim0.1$~keV thermal spectrum was expected, the observed spectra are non thermal and extend beyond 10~keV.
Some authors \citep{EliNatur,Waxman07,Alicia08D} argued that this is due to shock breakout physics.
Others argued that the X-ray bursts can not be explained within this framework and imply the existence of relativistic energetic jets penetrating through the stellar mantle/envelope \citep{Soderberg06,Fan06,Ghisellini07,Li07,Mazzali08,Li08}.
A recent derivation of the structure of mildly and highly relativistic radiation-mediated shocks \citep{Katz10} shows that fast, $v_s/c\gtrsim0.2$, radiation mediated shocks produce photons of energy far exceeding the $\sim0.1$~keV downstream temperature, reaching 10's to 100's of keV.
This suggests that the observed outbursts may indeed be due to shock breakout (Note, that for SN2006aj there is an additional challenge, beyond the X-ray spectrum: The energy inferred to be deposited in the fastest part of the ejecta far exceeds that expected from shock acceleration in the envelope).

Our focus in the current paper is not on the X-ray outburst, but rather on the early, $\sim1$~day, UV/O emission that follows it.
Model predictions for the early UV/O emission were derived mainly using numerical calculations \citep[e.g.][]{Falk78,Ensman92,Blinnikov00,Blinnikov02,Gezari08}.
Following the detection of SN2006aj, an analytic model has been constructed \citep{Waxman07}.
One of the main advantages of the analytic model is that it provides explicit analytic expressions for the dependence of the emission on model parameters, thus making both the use of observations for determining parameters and the identification of model uncertainties much easier and more straightforward.
It was shown, in particular, that the photospheric temperature of the expanding envelope depends mainly on the progenitor's radius and on the opacity, $T_{\text{ph}}$ approximately proportional to $R_*^{1/4}$, and that the luminosity $L$ is approximately proportional to $(E/M)R_*$.
This implies that the progenitor radius, which is only poorly constrained by other observations, and $E/M$ may be directly determined by measuring and analyzing the early UV/O emission.

In should be noted here that most numerical models, which are used for analyzing SN light curves, focus on the long term radioactively driven emission. Such models do not describe the early UV/O emission \citep[e.g. figure 3 in][]{Tanaka09}, largely due to the fact that they lack the resolution required to properly describe the evolution of the outer $\sim10^{-2}M_\odot$ part of the shock-heated envelope, which drives the early emission \citep{Waxman07}. Numerical calculations that allow a proper treatment of the early emission are, on the other hand, computationally very demanding \citep[see e.g.][]{Gezari08,Blinnikov00}. It is difficult to use these models for obtaining the dependence of predictions on model parameters and therefore for using observations to constrain these parameters.

In the near future, we expect an increasing rate of detection of early UV/O SN emission.
Ground based SN surveys with high rate sampling
\citep[less or order of a day, e.g.][]{Law09,quimby_phd} will detect SNe at the early stages of their expansion, with a bias towards detection of the more abundant type II SNe.
The detection of X-ray outbursts, that were observed to mark the onset of several SN Ib/c explosions, suggest that the early UV/O emission from these SNe may be detectable by follow-ups of X-ray triggers.
The MAXI (Monitor of All-sky X-ray Image) experiment on board the Kibo module \citep{Matsuoka97}, which was launched this year, is expected to detect events similar to SN2008D at a rate of up to a few per year.
The EXIST satellite \citep{Grindlay03,Band08}, which is still in planning, will further increase the event rate and improve the X-ray spectral coverage \citep[the detection rate may also be enhanced by smaller dedicated X-ray observatories, see e.g.][]{Calzavara04}.

The main goal of the current paper is to extend the analytic model \citep[][]{Waxman07} to include a more realistic description of the opacity and its variation with time (mainly due to recombination), and to include an approximate description of the deviation of the emitted spectrum from a black body spectrum (due to photon diffusion). Our results will facilitate the use of upcoming observations of the early emission from SNe for constraining progenitor and explosion parameters. We consider several types of progenitor envelopes: Dominated by Hydrogen, as appropriate for RSG and BSG progenitors, as well as envelopes dominated by He, C/O, and He-C/O mixtures representing different degrees of H/He stripping, due to wind mass loss \citep[e.g.][]{Woosley93} or binary interaction \citep[e.g.][]{Nomoto95}, and different evolution scenarios \citep[e.g. due to rotation induced mixing,][]{Meynet03,Crowther07}.

For completeness, we first present in \S~\ref{sec:const_opacity_model} the simple model derived in \citep{Waxman07}. The main assumptions adopted are that the envelope density drops near the stellar edge as a power of the distance from the edge, eq.~(\ref{eq:init_dnsty_prfl}), that the SN shock velocity may be approximated \citep[following][]{MM} by an interpolation between the Sedov--von Neumann--Taylor and the Gandel'Man-Frank-Kamenetskii--Sakurai self-similar solutions, eq.~(\ref{eq:vs_nonrel0}), and that the opacity, $\kappa$, is space and time independent. We also show, in \S~\ref{ssec:chck_adiabatic}, that the effects of photon diffusion on the predicted luminosity \citep[considered in][]{Chev,Chev08} are small. The model is extended in \S~\ref{sec:real_model} to include a more realistic description of the opacity and an approximate description of the effect of photon diffusion on the spectrum. In \S~\ref{sec:extinction} we show that the relative extinction at different wavelengths ($A_\lambda-A_V$) may be inferred from the light-curves at these wave-lengths. In \S~\ref{ssec:Applic_SG} we compare our model predictions to observations of the early emission available for two SNe, arising from RSG and BSG progenitors, and to detailed numerical simulations that were constructed to reproduce these observations. In \S~\ref{ssec:Applic_08D} we use our model to analyze the early UV/O observations of SN2008D. Our main results are summarized and discussed in \S~\ref{sec:Conclusion}.

\section{A simple model}
\label{sec:const_opacity_model}

We first derive in \S~\ref{ssec:non-rel_profile} the density, velocity and temperature profiles of the (post-breakout) expanding stellar envelope. We then derive in \S~\ref{ssec:prop_photo} the radius and temperature of the photosphere. These derivations are carried out under the simplifying assumption, that photon diffusion is negligible (and hence that the flow is adiabatic) below the photosphere. This assumption is justified in \S~\ref{ssec:chck_adiabatic}.

\subsection{Expanding ejecta profiles}
\label{ssec:non-rel_profile}

The UV/O emission on a day time scale arises from the outer $\lesssim10^{-2}M_\odot$ shell of the ejecta \citep{Waxman07}. Neglecting the shell's self-gravity and its thickness (relative to $R_*$), the pre-explosion density profile within the shell may be approximated by \citep{Chandrasekhar39}
\be
    \label{eq:init_dnsty_prfl}
    \rho_0(r_0) = \rho_{1/2} \delta^n,
\ee
where $ \delta \equiv (1- r_0/R_*)$, $r_0$ is the radius, and $n=3,3/2$ for radiative and efficiently convective envelopes respectively. \citet{MM} have shown that the velocity of the SN shock within the envelope is well approximated by an interpolation between the Sedov--von Neumann--Taylor and the Gandel'Man-Frank-Kamenetskii--Sakurai self similar solutions, \citep{VonNeumann47,Sedov59,Taylor50,GandelMan56,Sakurai60}
\be
    \label{eq:vs_nonrel0}
    v_s(r_0) = A_v \left[ \f{E}{m(r_0)} \right]^{1/2} \left[ \f{m(r_0)}{\rho r_0^3} \right]^{\beta_1},
\ee
where $E$ is the energy deposited in the ejecta, $m(r_0)$ is the ejecta mass enclosed within $r_0$, $A_v \simeq 0.79$ and $\beta_1 \simeq 0.19$. For $r_0\rightarrow R_*$, eq.~(\ref{eq:vs_nonrel0}) reduces to
\be
    \label{eq:vs_nonrel}
    v_s \simeq A_v \left( \f{E}{M} \right)^{1/2}
    \left( \f{4\pi}{3f_{\rho}} \right)^{\beta_1} \delta^{-\beta_1 n},
\ee
where $f_{\rho} \equiv \rho_{1/2}/ \overline{\rho}_0$ and $\overline{\rho}_0$ is the average ejecta density, $\overline{\rho}_0\equiv 3M/4\pi R^3_*$.
\citet[][Appendix A]{Calzavara04} have derived the values of $f_{\rho}$ expected for various progenitors (using their notation, $f_{\rho}=(3/4\pi)\rho_1/\rho_*$). For BSGs they find $f_{\rho}$ varying nearly linearly with mass,
from 0.031 to 0.062 for $8.5<M/M_{\odot}<18.5$, and for RSGs they find $ 0.079\lesssim  f_{\rho} \lesssim 0.13$.
We show below that the results are not very sensitive to the value of $f_{\rho}$.

In what follows we replace the Lagrangian coordinate $r_0$ with
\be
\label{eq:dmAsdl}
    \delta_m(\delta) \equiv M^{-1}\int^{R_*}_{(1-\delta)R_*} \text{d}r 4 \pi r^2 \rho_0(r) \simeq \f{3 f_{\rho}}{n+1}
    \delta^{n+1},
\ee
the fraction of the ejecta mass lying initially above $r_0$.

As the radiation mediated shock passes through a fluid element lying at $r_0$, it increases its pressure to
\begin{equation}\label{eq:p0}
    p_0 = \f{6}{7} \rho_0 v_s^2,
\end{equation}
and its density to $7\rho_0$ (recall that the post shock energy density is dominated by radiation). As the shocked fluid expands, it accelerates, converting its internal energy to kinetic energy.
\citet{MM} have shown that the final velocity, $v_f(r_0)$, of the fluid initially lying at $r_0$ is well approximated by $v_f(r_0)=f_v(r_0) v_s(r_0)$ with $f_v\approx2$.
The value of $f_v$ depends on the curvature of the shells.
The effect of this dependence is considered below.

Equations~(\ref{eq:vs_nonrel}) and~(\ref{eq:p0}) hold as long as the shock width is much smaller than the width of the stellar envelope shell lying ahead of the shock.
When the shock reaches a radius at which the optical depth of the shock transition layer, $\tau_s \cong c/v_s$, becomes comparable to the optical depth of the shell lying ahead of the shock, $\tau_0 \backsimeq M \delta_m \kappa/ 4\pi R_*^2$, the radiation "escapes" ahead of the shock, producing a "shock breakout flash", and the shock can no longer be sustained by radiation.
The mass fraction at which breakout takes place is \citep[e.g.][]{MM}
\be \label{eq:dmSB}
    \dmsb \backsimeq 2 \times 10^{-5}
    \f{f_{\rho}^{-0.07} R_{*,13}^{2.3}}
    {E_{51}^{0.57} (M/M_{\odot})^{0.57} \kappa^{1.1}_{0.34} }
\ee
For smaller values of $\delta_m$, $\delta_m<\dmsb$, the velocity and pressure profiles are shallower than given by eqs.~(\ref{eq:vs_nonrel}) and~(\ref{eq:p0}).

Assuming adiabatic expansion, neglecting photon diffusion below the photosphere, the pressure and density of the expanding fluid are related by
\be
    \label{eq:p_aftr_expns}
    \begin{split}
    p(\delta_m,t) = \bigg[ \f{\rho(\delta_m,t)}{7\rho_0(\delta_m)} \bigg]^{4/3} p_0(\delta_m).
    \end{split}
\ee
Once a fluid shell expands to a radius significantly larger than $R_*$, its pressure drops well below $p_0$ and its velocity approaches the final velocity $v_f$.
At this stage, $v_f t \gg R_* $, the shell's radius is approximately given by
\be
    \label{eq:r_prop_nonrel}
    r(\delta_m,t) \cong v_f(\delta_m) t
\ee
and its density is then given by
\be
    \begin{split}
    \rho = -\f{M}{4\pi r^2 t  }
    \left( \f{dv_f}{d\delta_m} \right)^{-1} \cong
    \f{n+1}{\beta_1\,n}  \, \f{ M}{4  \pi t^3 v_f^3 }
     \delta_m
    \end{split}.
\ee
The resulting density profile is steep, $d\ln\rho/d\ln r=d\ln\rho/d\ln v_f=-3-(n+1)/\beta_1n\approx-10$.

\subsection{Photospheric temperature and radius}
\label{ssec:prop_photo}

For a time and space independent opacity $\kappa$ (which applies, e.g., for opacity dominated by Thomson scattering with constant ionization), the optical depth of the plasma lying above the shell marked by $\delta_m$ is
\begin{eqnarray}
    \label{eq:optcl_dpth_nonrel}
    \tau(\delta_m,t) & \equiv & \int_{r(\delta_m,t)}^{\infty} \text{d}r \kappa \rho(r,t)
        = \f{\kappa M }{4\pi} \int_{0}^{\delta_m} \f{\text{d}\delta_m'}{r^2(\delta_m')} \nonumber
        \\
        & = & \f{1}{1+2 \beta_1 n/(1+n)} \f{\kappa M \delta_m}{4\pi t^2 v_f^2(\delta_m)},
\end{eqnarray}
where the last equality holds when eq.~(\ref{eq:r_prop_nonrel}) is satisfied.
We define the Lagrangian location of the photosphere, $\delta_{m,ph}$, by $\tau(\delta_m = \delta_{m,ph} ,t) = 1$.
We consider two type of envelopes: radiative envelopes typical to blue super giants (BSG), for which we take $n = 3$, and efficiently convective envelopes typical to red super giants (RSG), for which we take $n=3/2$ (note, that inefficient convection may lead to a more complicated density profile).
For  $n = 3/2$ and $n = 3$ envelopes we have
\be
    \label{eq:ph_prop_nonrel}
    \begin{split}
    \delta_{m,ph}(t) &= 2.4 \times 10^{-3} f_{\rho}^{-0.12}
    \f{ E_{51}^{0.81}}{(M/M_{\odot})^{1.6} \kappa^{0.81}_{0.34} }
    t_5^{1.63} \, (n=\f{3}{2}),\\
    \delta_{m,ph}(t) &= 2.6 \times 10^{-3} f_{\rho}^{-0.073}
    \f{ E_{51}^{0.78}}{(M/M_{\odot})^{1.6} \kappa^{0.78}_{0.34} }
    t_5^{1.56} \, (n=3),
    \end{split}
\ee
where $E = 10^{51}E_{51}\erg$, $t = 10^5 t_5\se$, and $\kappa =0.34 \kappa_{0.34} \cm^2/\gr$.
Here, and in what follows, we use \citep[following][]{MM} $ \beta_1 = 0.1909, f_v = 2.1649, \text{and }A_v = 0.7921 $ for $n=3/2$ and
$\beta_1 = 0.1858, f_v = 2.0351, \text{and }A_v = 0.8046 $ for $n=3$. Using eqs.~(\ref{eq:vs_nonrel}),~(\ref{eq:r_prop_nonrel}), and~(\ref{eq:p_aftr_expns}), we find that the radius and the effective temperature of the photosphere are given by
\be
\label{eq:non_rel_r_ph}
\begin{split}
    r_{\text{ph}}(t) &= 3.3 \times 10^{14} f_{\rho}^{-0.062}
    \f{E_{51}^{0.41} \kappa_{0.34}^{0.093} }{(M/M_{\odot})^{0.31}}
    t_5^{0.81}\cm \, (n=\f{3}{2}),\\
    r_{\text{ph}}(t) &= 3.3 \times 10^{14} f_{\rho}^{-0.036}
    \f{E_{51}^{0.39} \kappa_{0.34}^{0.11} }{(M/M_{\odot})^{0.28}}
    t_5^{0.78}\cm \, (n=3),\\
\end{split}
\ee
and
\be
    \label{eq:non_rel_T_ph}
    \begin{split}
    T_{\text{ph}}(t) =1.6 \, f_{\rho}^{-0.037}
    \f{E_{51}^{0.027} R_{*,13}^{1/4} }{(M/M_{\odot})^{0.054} \kappa^{0.28}_{0.34}}
    t_5^{-0.45} \eV \, (n=\f{3}{2}),\\
    T_{\text{ph}}(t) =1.6 \, f_{\rho}^{-0.022}
    \f{E_{51}^{0.016} R_{*,13}^{1/4} }{(M/M_{\odot})^{0.033} \kappa^{0.27}_{0.34}}
    t_5^{-0.47} \eV \, (n=3).\\
    \end{split}
\ee
Here, $R_* = 10^{13} R_{*,13} \cm $. The dependence on $n$ and on $f_{\rho}$ is weak. Note, that eq.~(\ref{eq:non_rel_T_ph}) corrects a typo (in the numerical coefficient) in eq.~(19) of \citep{Waxman07}.

As mentioned in the introduction, the photospheric temperature is weakly dependent on $E$ and $M$ and is approximately linear in $(R_*/\kappa)^{1/4}$.
The photospheric radius, on the other hand, does not depend on $R_*$, is weakly dependent on $\kappa$ and is approximately linear in $E^{0.4}/M^{0.3}$.
The luminosity predicted by the simple model described here, $L = 4\pi \sigma r_{\text{ph}}^2 T_{\text{ph}}^4$, is
\be
    \label{eq:L_RSG}
    L = 8.5\times 10^{42} \f{E^{0.92}_{51} R_{*,13} }
    {f^{0.27}_{\rho} (M/M_{\odot})^{0.84}\kappa^{0.92}_{0.34} }t^{-0.16}_5 \erg\, {\rm s}^{-1}
\ee
for $n=3/2$, and
\be
    \label{eq:L_BSG}
    L = 9.9\times 10^{42} \f{E^{0.85}_{51} R_{*,13} }
    {f^{0.16}_{\rho} (M/M_{\odot})^{0.69}\kappa^{0.85}_{0.34} }t^{-0.35}_5 \erg\, {\rm s}^{-1}
\ee
for $n=3$.

Our simple description of $r_{\text{ph}}$ $T_{\text{ph}}$, and $L$, eqs.~(\ref{eq:non_rel_r_ph}--\ref{eq:L_BSG}), holds for $\delta_{m,\rm ph}>\dmsb$, i.e. for (comparing eqs.~(\ref{eq:dmSB}) and~(\ref{eq:ph_prop_nonrel}))
\be
    \label{eq:t_bo}
    t>t_{\rm BO}=0.05\times10^5 \f{(M/M_{\odot})^{0.6} R_{*,13}^{1.4} }{\kappa^{0.2}_{0.34}E^{0.9}_{51} }\rm s
\ee
This requirement also ensures that the ejecta shells have expanded and cooled significantly, and thus reached their terminal velocity $v_f $. For these times, the approximation of eq.~\eqref{eq:p_aftr_expns} holds, and the radius of each shell is well approximated by $r = t \, v_f $.

The value of $f_v$ deviates from 2 for large $\delta_m$, due to the increasing curvature of the shells \citep[e.g.][]{MM}.
Requiring the deviation not to exceed $30\%$, in which case the error in eq.~(\ref{eq:non_rel_T_ph}) is smaller than $15\%$, implies limiting the analysis to times
\begin{equation}
\label{eq:t_vf2}
    t  < (f_{\rho}/0.07)^{0.69} (M/M_{\odot})
    \kappa_{0.34}^{0.5}  E_{51}^{-0.5} \times t_{\rm fv} ,
\end{equation}
where $t_{\rm fv}  = 4.5 \times 10^6{\rm s} , 1.2 \times 10^5 \rm s$ for $n = 3/2$ and $n = 3$ envelopes respectively.
Thus, the curvature effect is negligible on day-week time scale for $M \sim 10 M_{\odot}$, and may be significant on day time scale only for low mass ejecta.


\subsection{Photon diffusion}
\label{ssec:chck_adiabatic}

Let us next examine the assumption, that photon diffusion does not lead to strong deviations from adiabatic expansion below the photosphere. The size of a region around $r(\delta_m,t)$ over which the diffusion has a significant effect is $D(\delta_m,t) \simeq \sqrt{ct/3k\rho(\delta_m,t)}$. Thus, the radius $r_d=r(\delta_{m,d},t)$ above which diffusion affects the flow significantly may be estimated as $D(\delta_m=\delta_{m,d},t)=r(\delta_m=\delta_{m,d},t)$. This gives
\be
\begin{split}
\label{eq:r_d}
    r_d(t) &= 3.7 \times 10^{14} f_{\rho}^{-0.069}
    \f{E_{51}^{0.45} \kappa_{0.34}^{0.1} }{(M/M_{\odot})^{0.35}}
    t_5^{0.79}\cm \, (n=\f{3}{2}),\\
    r_d(t) &= 3.8 \times 10^{14} f_{\rho}^{-0.04}
    \f{E_{51}^{0.44} \kappa_{0.34}^{0.12} }{(M/M_{\odot})^{0.32}}
    t_5^{0.75}\cm \, (n=3).\\
\end{split}
\ee
This radius is similar, and somewhat larger than, the photospheric radius given by Eq.~(\ref{eq:non_rel_r_ph}). The rapid increase of the diffusion time, $\sim 3\kappa\rho r^2/c$, at smaller radii implies that diffusion does not significantly affect the fluid energy density below the photosphere.
Next, we note that in regions where the diffusion time is short, the luminosity carried by radiation, $L\propto r^2 dp/d\tau$, is expected to be independent of radius.
The steep dependence of the density on radius ($d\ln\rho/d\ln r\sim-10$) then implies that the energy density roughly follows $p\propto\tau$, i.e. $T\propto\tau^{1/4}$. This temperature profile is close to the adiabatic profiles derived in \S~\ref{ssec:non-rel_profile}, for which $T\propto \tau^{0.28}, \tau^{0.27}$ for $n=3,3/2$ respectively. Thus, diffusion does not lead to a significant modification of the pressure and temperature profiles also at radii where the diffusion time is short.

The validity of the above conclusions may be tested by using the self-similar solutions of \citet{Chev}, which describe the diffusion of photons in an expanding envelope with a density following $\rho\propto r^{-m}t^{m-3}$ and initial pressure $p\propto r^{-l}t^{l-4}$. The evolution of the ejecta density and pressure derived in \S~\ref{ssec:non-rel_profile} follows, for $v_f t\gg R_*$, $\rho=B r^{-m}t^{m-3}$ with $m-3=(1+n)/n\beta_1$, and $p=A r^{-l}t^{l-4}$ with $l=(3\gamma-2)+(\gamma+n)/n\beta_1$, where $\gamma = 4/3$. Applying the solutions of \citet{Chev} to these profiles one finds the same $r_d$ as given by eq.~(\ref{eq:r_d}) and
\be
    L_c = 9.6\times 10^{42} \f{E^{0.91}_{51} R_{*,13} }
    {f^{0.17}_{\rho} (M/M_{\odot})^{0.74}\kappa^{0.82}_{0.34} }t^{-0.35}_5 \erg\, {\rm s}^{-1},
\ee
for $n=3$ and
\be
    L_c = 1.0\times 10^{43} \f{E^{0.96}_{51} R_{*,13} }
    {f^{0.28}_{\rho} (M/M_{\odot})^{0.87}\kappa^{0.91}_{0.34} }t^{-0.17}_5 \erg\, {\rm s}^{-1},
\ee
for $n=3/2$ (The parameter $q$ of the self-similar solutions is $q=0.495, 0.683$ for $n=3$ and $n=3/2$ respectively; The density and pressure coefficients are
$A= 2.95 E^{3.89} R_{*} f^{-0.44} M^{-2.89},
53.7 E^{4.95} R_{*} f^{-0.89} M^{-3.95}$ and
$B=100 E^{3.59} f^{-0.33} M^{-2.59},
10^3 E^{4.37} f^{-0.67} M^{-3.37}$
for $n=3$ and $n=3/2$ respectively).

The parameter dependence of $L_c$ derived using the self-similar diffusion solutions is similar to that obtained by the simple model of \S~\ref{ssec:prop_photo}, and the normalization of $L_c$ derived using the self-similar diffusion solutions differs from the results of \S~\ref{ssec:prop_photo}, eqs.~(\ref{eq:L_BSG}) and~(\ref{eq:L_RSG}), by $\approx10\%$. The effective temperatures derived from the diffusion solutions, via $L=4\pi r_{\text{ph}}^2\sigma T^4$, differ from those derived in the previous section by 1\% to 5\%.

Thus, the effects of diffusion on the luminosity and on the effective temperature are small, as expected. It is important to emphasize in this context that since the diffusion approximation breaks down near the photosphere, the results obtained using the self-similar diffusion solutions are not necessarily more accurate than those derived (e.g. in the previous section) by neglecting photon diffusion below the photosphere. Improving the accuracy of the simple model requires a transport, rather than a diffusion, description of the photon propagation. The differences between the results obtained neglecting diffusion and including it may be considered as a rough estimate of the inaccuracy of the model.

Finally, the following note is in place here. The results of \citet{Chev} for $L$ and $r_d$ (eqs. 3.19 and 3.20) are different both in normalization and in scaling from those derived here. This is due to some typographical errors in earlier eqs. of that paper. When corrected, in \citet{Chev08}, the results obtained using the diffusion solutions are similar to those obtained in \citet{Waxman07} and here. The difference in the numerical coefficient of the photospheric temperature, eq. (19) of \citet{Waxman07} and eq. (5) of  \citet{Chev08}, is due mainly to the typo in  eq.~(19) of \citet{Waxman07}, which is corrected in eq.~(\ref{eq:non_rel_T_ph}) above.

\section{Model extensions}
\label{sec:real_model}

The approximation of space and time independent opacity is justified at early times, when the envelope is highly ionized and the opacity is dominated by Thomson scattering. On a day time scale, the temperature of the expanding envelope drops to $\sim1$~eV, see eq.~(\ref{eq:non_rel_T_ph}). At this temperature, significant recombination may take place, especially for He dominated envelopes, leading to a significant modification of the opacity. The model presented in \S~\ref{sec:const_opacity_model} is generalized in \S~\ref{ssec:var_opacity} to include a more realistic description of the opacity. The deviation of the emitted spectrum from a black body spectrum, due to photon diffusion, is discussed in \S~\ref{ssec:calc_T_col}. A brief discussion of the effect of line opacity enhancement due to velocity gradients is given in \S~\ref{ssec:exp_opacity}.
Throughout this section, we use the density structure given by eq.~(\ref{eq:init_dnsty_prfl}) with $n=3$, as appropriate for radiative envelopes. As explained in the previous section, the results are not sensitive to the exact value of $n$.

\subsection{Varying opacity}
\label{ssec:var_opacity}

In order to obtain a more accurate description of the early UV/O emission, we use the mean opacity provided in the OP project tables \citep{OPCD}. We replace eq.~(\ref{eq:optcl_dpth_nonrel}) with
\be
\label{eq:tau_exact}
    \tau(\delta_{m},t)=
    \int_{r(\delta_{m},t)}^{\infty}\text{d}r\rho\,\kappa[T(\delta_{m},t),\rho(\delta_{m},t)],
\ee
where $\kappa(T,\rho)$ is the Rosseland mean of the opacity, and solve $\tau(\delta_m = \delta_{m,ph} ,t) = 1$ numerically for the location of the photosphere.
In order to simplify the comparisons with the suggested analytical models, in the reminder of this section we shall take the ejecta properties in the limit of eq.~(\ref{eq:r_prop_nonrel}).

\subsubsection{H envelopes}
\label{ssec:H}

\begin{figure}
\includegraphics[scale=1]{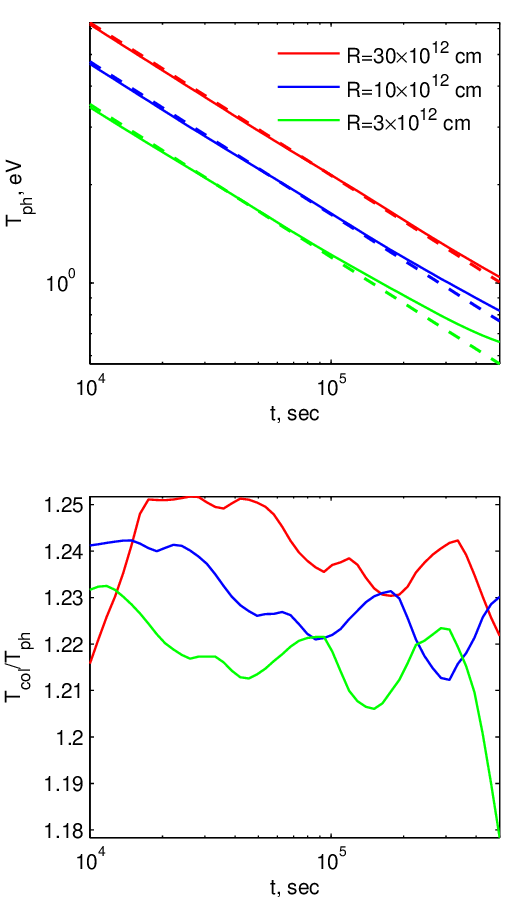}
\caption{Photospheric temperature (top panel) and the ratio of color to effective (=photospheric) temperatures (bottom panel) for explosions in H dominated envelopes (H with solar composition of heavier elements). The top panel compares the analytic approximation of eq.~(\ref{eq:non_rel_T_ph}) for fixed, $\kappa=0.34{\rm cm^2/g}$, opacity (dashed lines) with the numerical solution of eq.~(\ref{eq:tau_exact}) using OP table opacities (solid lines). The color to effective temperature ratio was calculated as explained in \S~\ref{ssec:calc_T_col}. Results are shown for $E=10^{51}$~erg, $M=1\, M_{\odot}$, and three progenitor radii, $\{30, 10, 3\}\times 10^{12} \cm$.
\label{fig:EffModelHNew}
}
\end{figure}

Consider first explosions in H dominated envelopes. In figure~\ref{fig:EffModelHNew} we compare the temperature of the photosphere calculated using the OP tables with the results given by eq.~(\ref{eq:non_rel_T_ph}) for $\kappa=0.34{\rm cm^2/g}$, corresponding to fully ionized H. The difference in $T_{\text{ph}}$ obtained by the two methods is smaller than 10\% for $T_{\text{ph}}>1$~eV. At lower temperatures, the $\kappa=0.34{\rm cm^2/g}$ approximation leads to an underestimate of $T_{\text{ph}}$, by $\approx20\%$ at 0.7~eV. This is due to the reduction in opacity accompanying H recombination. The reduced opacity implies that the photosphere penetrates deeper into the expanding envelope, to a region of higher temperature. The photospheric radius is not significantly affected and is well described by eq.~(\ref{eq:non_rel_r_ph}).

\subsubsection{He envelopes}
\label{ssec:He}

Let us consider next explosions in He dominated envelopes. In this case, the constant opacity approximation does not provide an accurate approximation for $T_{\text{ph}}$. We therefore replace eqs.~(\ref{eq:non_rel_r_ph}) and~(\ref{eq:non_rel_T_ph}) with an approximation which takes into account the reduction of the opacity due to recombination. On the time scale of interest, hour~$\lesssim t \lesssim $~day, the photospheric temperature is in the energy range of 3eV~$\gtrsim T \gtrsim$~1eV. In this temperature range (and for the characteristic densities of the photosphere), the opacity may be crudely approximated by a broken power law,
\be
\label{eq:approxHe_opacity}
\kappa = 0.085 \,\OpUnit
\begin{cases}
     (T/\TbHe)^{0.88}  & T > \TbHe \\
     (T/\TbHe)^{10}    & T \leq \TbHe
\end{cases}\,.
\ee
Using this opacity approximation, we find that eq.~(\ref{eq:non_rel_T_ph}) for the photospheric temperature is modified to
\be
\label{eq:EffModel-THe}
T_{\text{ph}}(t) =
\begin{cases}
    1.33 \eV f_\rho^{-0.02}R_{*,12}^{0.20} t_5^{-0.38} & T_{\text{ph}} \geq \TbHe \\
    \TbHe (t/t_b)^{-0.12}                         &  T_{\text{ph}} < \TbHe
\end{cases}.
\ee
Here, $R_*=10^{12}R_{*,12}$~cm and $t_b$ is the time at which $T_{\text{ph}}=\TbHe$, and we have neglected the dependence on $E$ and $M$, which is very weak. The photospheric radius, which is less sensitive to the opacity modification, is approximately given by
\be
\label{eq:EffModel-r}
    r_{\text{ph}}(t) = 2.8\times 10^{14} f_\rho^{-0.038} E_{51}^{0.39} (M/M_{\odot})^{-0.28} t_5^{0.75} \cm.
\ee
Here we have neglected the dependence on $R_*$, which is weak. For $T_{\text{ph}}>\TbHe$, the bolometric luminosity is given by
\be
    \label{eq:L_He}
    L = 3.3\times 10^{42}
    \f{E^{0.84}_{51} R_{*,12}^{0.85} }
    {f^{0.15}_{\rho} (M/M_{\odot})^{0.67} }t^{-0.03}_5 \erg\, {\rm s}^{-1}.
\ee
The deviation of $f_v$ from the assumed value of 2 for large $\delta_m$ (discussed in \sref{ssec:prop_photo}) has only a negligible effect on eqs.~\eqref{eq:EffModel-THe} and~\eqref{eq:L_He}.

\begin{figure}
\includegraphics[scale=1]{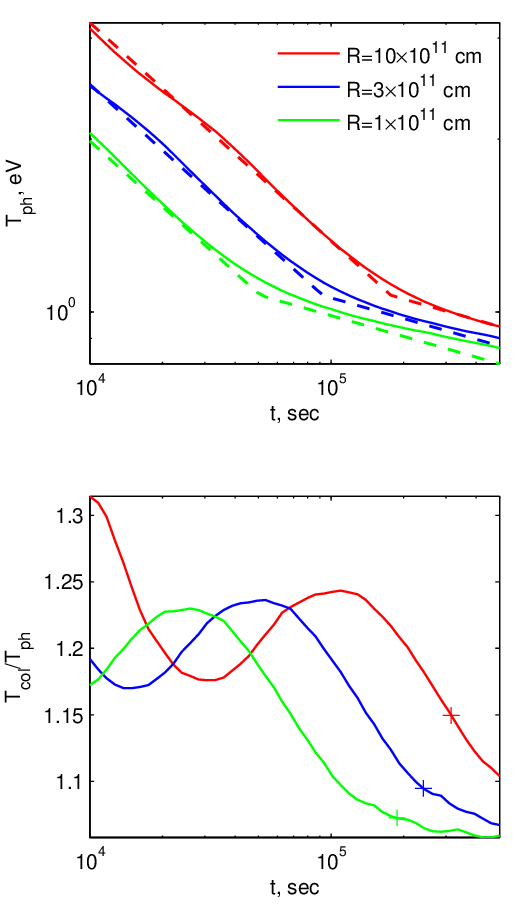}
\caption{
Photospheric temperature (top panel) and the ratio of color to effective (=photospheric) temperatures (bottom panel) for explosions in He dominated envelopes \citep[Helium mass fraction of 98\%, C/N/O/Ne fractions taken from fig.~18 of][]{Meynet03}. The top panel compares the analytic approximation of eq.~(\ref{eq:EffModel-THe}) (dashed lines) with the numerical solution of eq.~(\ref{eq:tau_exact}) using OP table opacities (solid lines). The color to effective temperature ratio was calculated as explained in \S~\ref{ssec:calc_T_col}. Results are shown for $E=10^{51}$~erg, $M=1\, M_{\odot}$, and three progenitor radii, $10, 3, 1\times 10^{11} \cm$. The + sign indicates the time at which $\delta m_{\text{ph}}=\dmbreak $. \label{fig:EffModelHe-WRNew}}
\end{figure}

In figure~\ref{fig:EffModelHe-WRNew} we compare the approximation of eq.~(\ref{eq:EffModel-THe}) for $T_{\text{ph}}$ with a numerical calculation using the OP opacity tables. The approximation of eq.~(\ref{eq:EffModel-THe}) holds to better than 8\% down to $T_{\text{ph}} \simeq 1$~eV. The temperature does not decrease significantly below $\simeq 1$~eV due to the rapid decrease in opacity below this temperature, which is caused by the nearly complete recombination.

The following comment is in place here. The strong reduction in opacity due to He recombination implies that the photosphere reaches deeper into the envelope, to larger values of $\delta_m$. The plus signs in fig.~\ref{fig:EffModelHe-WRNew} denote the time at which $\delta_{m,ph}=0.1$. For such a large mass fraction, the initial density profile is no longer described by eq.~(\ref{eq:init_dnsty_prfl}) and the evolution of the ejecta is no longer given by the eqs. of \S~\ref{ssec:non-rel_profile}. This further complicates the model for the emission on these time scales (see \S~\ref{ssec:mod_ejecta} for further discussion).

\subsubsection{He-C/O envelopes}
\label{ssec:HeCO}

Let us consider next envelopes composed of a mixture of He and C/O. At the relevant temperature and density ranges, the C/O opacity is dominated by Thomson scattering of free electrons provided by these atoms, and is not very sensitive to the C:O ratio. Denoting by 1-Z the He mass fraction, the C/O contribution to the opacity may be crudely approximated, within the relevant temperature and density ranges, by
\be
\label{eq:approxCO_opacity}
\kappa =  0.043\,\mbox{Z}\,\OpUnit (T/1\,\eV)^{1.27}.
\ee

This approximation holds for a 1:1 C:O ratio. However, since the opacity is not strongly dependent on this ratio, $T_{\text{ph}}$ obtained using eq.~(\ref{eq:approxCO_opacity}) (eq.~\ref{eq:EffModel-TCO}) holds for a wide range of C:O ratios (see discussion at the end of this subsection). At the regime where the opacity is dominated by C/O, eq.~(\ref{eq:non_rel_T_ph}) is modified to
\be
\label{eq:EffModel-TCO}
T_{\text{ph}}(t) = 1.5 \eV f_\rho^{-0.017} \mbox{Z}^{-0.2}  R_{*,12}^{0.19} t_5^{-0.35}.
\ee

\begin{figure}
\includegraphics[scale=1]{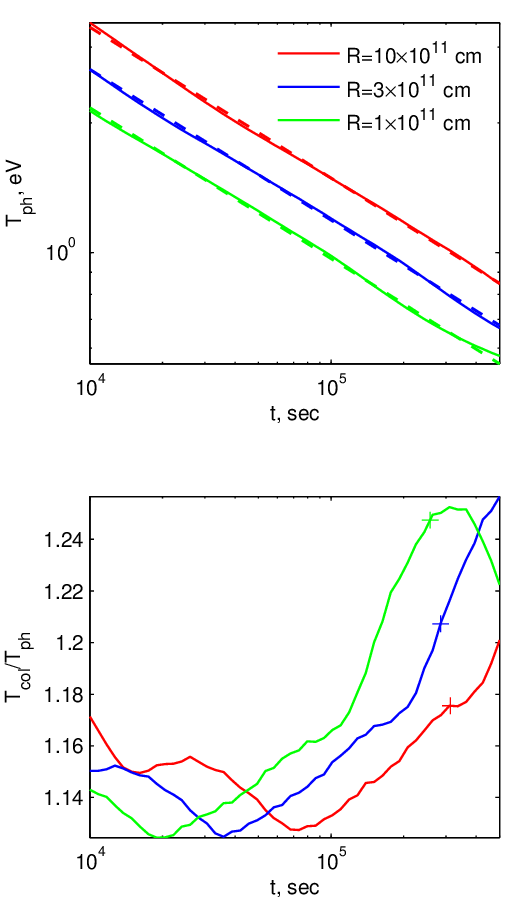}
\caption{Photospheric temperature (top panel) and the ratio of color to effective (=photospheric) temperatures (bottom panel) for explosions in a C/O envelopes (1:1 C:O ratio). The top panel compares the analytic approximation of eq.~(\ref{eq:EffModel-TCO}) (dashed lines) with the numerical solution of eq.~(\ref{eq:tau_exact}) using OP table opacities (solid lines). The color to effective temperature ratio was calculated as explained in \S~\ref{ssec:calc_T_col}. Results are shown for $E=10^{51}$~erg, $M=1\, M_{\odot}$, and three progenitor radii, $10, 3, 1\times 10^{11} \cm$. The + sign indicates the time at which $\delta m_{\text{ph}}=\dmbreak$.
\label{fig:EffModelWONew}
}
\end{figure}

\begin{figure}
\includegraphics[scale=1]{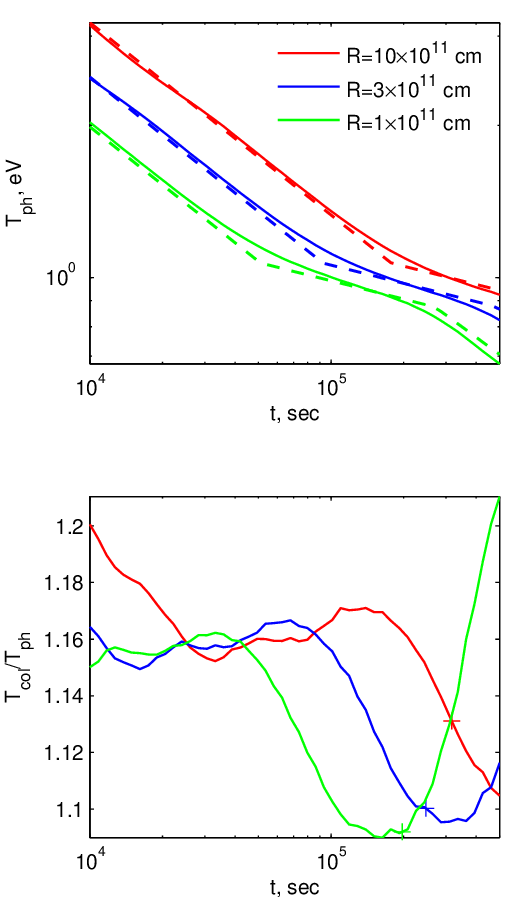}
\caption{Photospheric temperature (top panel) and the ratio of color to effective (=photospheric) temperatures (bottom panel) for explosions in a He-C/O envelopes (1-Z=0.7 He mass fraction, 2:1 C:O ratio). The top panel compares the analytic approximations obtained using eqs.~(\ref{eq:EffModel-THe}) and~(\ref{eq:EffModel-TCO}) with a transition temperature given by eq.~(\ref{eq:EffModel-TransT}) (dashed lines) with the numerical solutions of eq.~(\ref{eq:tau_exact}) using OP table opacities (solid lines). The color to effective temperature ratio was calculated as explained in \S~\ref{ssec:calc_T_col}. Results are shown for $E=10^{51}$~erg, $M=1\, M_{\odot}$, and three progenitor radii, $10, 3, 1\times 10^{11} \cm$. The + sign indicates the time at which $\delta m_{\text{ph}}=\dmbreak$.
\label{fig:EffModelHeCO-WR70New}
}
\end{figure}

In the absence of He, i.e. for $\mbox{Z}=1$, $T_{\text{ph}}$ is simply given by eq.~(\ref{eq:EffModel-TCO}). For a mixture of He-C/O, $\mbox{Z}<1$, $T_{\text{ph}}$ may be obtained as follows. At high temperature, where He is still ionized, the He and C/O opacities are not very different and $T_{\text{ph}}$ obtained for a He envelope, eq.~(\ref{eq:EffModel-THe}), is similar to that obtained for a C/O envelope, eq.~(\ref{eq:EffModel-TCO}). At such temperatures, we may use eq.~(\ref{eq:EffModel-THe}) for an envelope containing mostly He, and  eq.~(\ref{eq:EffModel-TCO}) with $\mbox{Z}=1$ for an envelope containing mostly C/O (a more accurate description of the Z-dependence may be straightforwardly obtained by an interpolation between the two equations). At lower temperature, the He recombines and the opacity is dominated by C/O. At these temperatures, $T_{\text{ph}}$ is given by eq.~(\ref{eq:EffModel-TCO}) with the appropriate value of $\mbox{Z}$. The transition temperature is given by
\be
\label{eq:EffModel-TransT}
    T_{\rm He-C/O} = 1\, \mbox{Z}^{0.1} \eV.
\ee
The photospheric radius, which is less sensitive to the opacity variations, is well approximated by eq.~(\ref{eq:EffModel-r}).
At the stage where the opacity is dominated by C/O, the bolometric luminosity is given by
\be
    \label{eq:L_CO}
    L = 4.7\times 10^{42} \f{E^{0.83}_{51} R_{*,12}^{0.8} }
    {f^{0.14}_{\rho} \mbox{Z}^{0.63}(M/M_{\odot})^{0.67} }t^{0.07}_5 \erg\, {\rm s}^{-1}.
\ee
The deviation of $f_v$ from the assumed value of 2 for large $\delta_m$ (discussed in \sref{ssec:prop_photo}) does not affect significantly the results of eqs.~\eqref{eq:EffModel-TCO} and~\eqref{eq:EffModel-TransT}, but may affect significantly the result given in eq.~\eqref{eq:L_CO}.
Eqs.~\eqref{eq:t_vf2}, \eqref{eq:approxCO_opacity} and~\eqref{eq:EffModel-TCO} indicate that for an explosion with $M = 10 M_{\odot}$ and $E_{51} =1$, the luminosity will be reduced by a factor of $\sim 2$ (compared to the predictions of eq.~\eqref{eq:L_CO}) when $T_{\text{ph}} \approx 1\eV$. This further complicates the model for the emission on these time scales. See \S~\ref{ssec:mod_ejecta} for further discussion.

In figures~\ref{fig:EffModelWONew} and~\ref{fig:EffModelHeCO-WR70New} we compare the analytic approximation for $T_{\text{ph}}$ derived above to the results of numerical calculations using the OP opacity tables. For the C/O envelopes (figure~\ref{fig:EffModelWONew}), the approximation of eq.~(\ref{eq:EffModel-TCO}) holds to better than 6\% down to $T_{\ph} \simeq0.5$~eV. For the Z=0.3 mixed He-C/O envelopes (figure~\ref{fig:EffModelHeCO-WR70New}), the approximations obtained by using eqs.~(\ref{eq:EffModel-THe}) and~(\ref{eq:EffModel-TCO}) with a transition temperature given by eq.~(\ref{eq:EffModel-TransT}) hold to better than 10\% down to $T_{\text{ph}} \simeq 0.8$~eV. Using similar comparisons for different compositions we find that similar accuracies are obtained over the range $0.7> \mbox{Z}>0.3$, and for increasing or decreasing the C:O ratio by an order of magnitude.

\subsection{Color vs. effective temperature}
\label{ssec:calc_T_col}

We have shown in \S~\ref{ssec:chck_adiabatic} that photon diffusion is not expected to significantly affect the luminosity. Such diffusion may, however, modify the spectrum of the emitted radiation. We discuss below in some detail the expected modification of the spectrum.

For the purpose of this discussion, it is useful to define the "thermalization depth", $r_{\text{ther}}$, and the "diffusion depth", $r_{\text{diff}}$. $r_{\text{ther}}(t)<r_{\text{ph}}(t)$ is defined as the radius at which photons that reach $r_{\text{ph}}(t)$ at $t$ "thermalize", i.e. the radius from which photons may reach the photosphere without being absorbed on the way. This radius may be estimated as the radius for which $\tau_{\text{sct}} \tau_{\text{abs}} \approx 1 $ \citep{Mihalas84}, where $\tau_{\text{sct}}$ and $\tau_{\text{abs}}$ are the optical depths for scattering and absorption provided by plasma lying at $r>r_{\text{ther}}(t)$. $r_{\text{ther}}$ is thus approximately given by
\be
\label{eq:r_thrm}
    3 (r_{\text{ther}}-r_{\text{ph}})^2 \kappa_{\text{sct}}(r_{\text{\text{ther}}}) \kappa_{\text{abs}}(r_{\text{\text{ther}}}) \rho^2(r_{\text{\text{ther}}}) = 1,
\ee
where $\kappa_{\text{\text{sct}}} $ and $\kappa_{\text{abs}}$ are the scattering and absorption opacities respectively (typically, the opacity is dominated by electron scattering). $r_{\text{diff}}$ is defined as the radius (below the photosphere) from which photons may escape (i.e. reach the photosphere) over a dynamical time (i.e. over $t$, the time scale for significant expansion). We approximate $r_{\text{diff}}$ by
\be
\label{eq:r_diff}
    r_{\text{ph}} = r_{\text{diff}}+\sqrt{c \, t/3 \kappa_{\text{\text{sct}}}(r_{\text{diff}}) \rho(r_{\text{diff}})},
\ee
where $c$ is the speed of light.

For $r_{\text{diff}} < r_{\text{\text{ther}}}$, photons of characteristic energy $3T(r_{\text{\text{ther}}},t)>3T_{\text{ph}}$ will reach the photosphere, while for $r_{\text{\text{ther}}} < r_{\text{diff}}$ photons of characteristic energy $3T(r_{\text{diff}},t)>3T_{\text{ph}}$ will reach the photosphere. Thus, the spectrum will be modified from a black body at $T_{\text{ph}}$ and its color temperature will be $T_{\text{col}}>T_{\text{ph}}$. We approximate in what follows $T_{\text{col}}=T(r_{\text{\text{ther}}})$ for $r_{\text{diff}} < r_{\text{\text{ther}}}$ and $T_{\text{col}}=T(r_{\text{diff}})$ for $r_{\text{diff}} > r_{\text{\text{ther}}}$.

The lower panels of figures \ref{fig:EffModelHNew}, \ref{fig:EffModelHe-WRNew}, \ref{fig:EffModelWONew} and \ref{fig:EffModelHeCO-WR70New} present the ratio $T_{\text{col}}/T_{\text{ph}}$ for the various envelope compositions considered. For this calculation, we have assumed that the scattering opacity is dominated by Thomson scattering of free electrons, used the electron density (as function of density and temperature) provided by the OP tables for determining $\kappa_{\text{\text{sct}}}$, and estimated $\kappa_{\text{abs}}=\kappa-\kappa_{\text{\text{sct}}}$ (recall that $\kappa$ is the Rosseland mean of the opacity). It would have been more accurate to use an average of the absorptive opacities over the relevant wavebands, which are not provided by the OP table. However, since the dependence of the color temperature on the absorptive opacity is weak, $T_{\text{col}} \propto \kappa_{\text{\text{abs}}}^{(-1/8)}$, the corrections are not expected to be large. The figures imply that over the relevant time scale, $t\lesssim1$~day,
\be
    \label{eq:f_T}
    f_T \equiv T_{\text{\text{col}}}/T_{\text{\text{ph}}} \approx 1.2\quad.
\ee

Using eq.~(\ref{eq:f_T}) with eqs.~(\ref{eq:non_rel_T_ph}),~(\ref{eq:EffModel-THe}) and~(\ref{eq:EffModel-TCO}) for the photospheric (effective) temperature, the progenitor radius may be approximately inferred from the color temperature by \be
    \label{eq:R_H}
    R_{*} \approx  0.70\times10^{12} \left[ \frac{T_{\text{col}}}{(f_T/1.2) \eV} \right]^{4}t_5^{1.9}
    f_{\rho}^{0.1}\,\rm cm
\ee
for H envelopes,
\be
    \label{eq:R_He}
    R_{*} \approx  1.2\times 10^{11} \left[ \frac{T_{\text{col}}}{(f_T/1.2) \eV} \right]^{4.9}   t_5^{1.9} f_{\rho}^{0.1}\, \cm
\ee
for He envelopes with $T>\TbHe$, and
\be
    \label{eq:R_He-CO}
    R_{*} \approx  0.58 \times 10^{11} \left[ \frac{T_{\text{col}}}{(f_T/1.2) \eV} \right]^{5.3}   t_5^{1.9} f_{\rho}^{0.1} \mbox{Z} \,\cm
\ee
for He-C/O envelopes when the C/O opacity dominates (the transition temperature is given in eq.~\eqref{eq:EffModel-TransT}).

\subsection{"Expansion opacity"}
\label{ssec:exp_opacity}

We have neglected in our analysis the effective broadening of atomic lines due to the velocity gradients in the outflow. Line broadening may have a significant effect on the opacity and on the observed emission \citep{karp77,Wagoner91,Eastman93}, as well as on the dynamics \cite[e.g. in the case of stellar winds, see][]{Friend83}. We give below a crude estimate of the line broadening effects for the problem of interest here. A detailed analysis of line broadening, which requires detailed numerical calculations \citep[see, e.g., Sec. 6.9 of ][]{Castor04}, is beyond the scope of the present manuscript.

The analysis of \citet{Wagoner91} shows that the effective line opacity introduced by the velocity gradients may significantly affect the Rosseland mean opacity (at the relevant densities and velocity gradients, see their figs.~5,~6) at temperatures where recombination leads to a large reduction of the Thomson electron scattering opacity.
The main contribution to this "expansion opacity" is from resonant line scattering of Fe group elements.
As explained in \S~\ref{ssec:var_opacity}, the main effect that recombination introduces to our analysis is the penetration of the photosphere to shells of high enough temperatures where significant ionization is maintained.
Since the opacity enhancement due to velocity gradient effects does not prevent the strong reduction of the opacity due to recombination, it will not prevent the penetration of the photosphere to a region of significant ionization. Nevertheless, the enhanced line opacity may introduce an order unity increase of the opacity at temperatures close to the recombination temperature at short, $\lambda<0.25\,\mu$, wavelengths \cite[see figs.~6,~7 of][]{Wagoner91}. A detailed analysis of this effect is beyond the scope of the current manuscript.

We should note, nevertheless, that the effective line opacity enhancement due to velocity gradients may be smaller in our case compared to the estimates of earlier analyses.
To show this, let us examine the following heuristic derivation of the effective broadening of atomic lines.
Consider a photon that travels outward/inward in the region of relatively low optical depth.
Due to the velocity gradient of the expanding ejecta, the photon frequency as measured in the plasma rest frame is shifted as it propagates by $d\nu/\nu=-dv/c=-(\partial v/\partial r)dr/c=-(\partial v/\partial r) dt$, where $dv=(\partial v/\partial r)dr$ is the velocity difference across $dr$ and we neglect the plasma speed with respect to that of the photon in setting $dr=c dt$.
Assuming that a photon is absorbed/scattered as its frequency is shifted across that of a line, then the probability for scattering/absorption is $dP=d\nu(dN/d\nu)$ where $dN/d\nu$ is the line "density" per unit frequency. The resulting photon mean free path is therefore $l^{-1}\sim dP/cdt \sim c^{-1}|\partial v/\partial r|\nu(dN/d\nu)$. For $v=r/t$, which is valid at late time, we have
\begin{equation}\label{eq:l}
    l_\nu^{-1}\sim\frac{|\partial v/\partial r|}{c}\nu(dN/d\nu)=(ct)^{-1}\nu(dN/d\nu)
\end{equation}
\citep[compare, e.g., to eq. (3.10) of][]{Wagoner91}. The photon is absorbed/scattered provided the line optical depth is large enough. Neglecting the natural width of the lines, we may replace the line opacity with $\kappa_\nu=\kappa_l\nu_0\delta(\nu-\nu_0)$. Denoting $\nu'=(1-dv/c)\nu$ we obtain
\begin{eqnarray}\label{eq:l_tau}
    \tau_l\sim\int cdt' \rho\kappa_{\nu'}&\approx&
    \int\frac{d\nu'}{\nu}\frac{c}{\partial v/\partial r}\rho\kappa_{\nu'}
    \nonumber \\ &\approx& \frac{c}{\partial v/\partial r}\rho\kappa_l=ct\rho\kappa_l.
\end{eqnarray}
Thus, the line should be "counted" in determining $dN/d\nu$ if $ct\rho\kappa_l\gg1$ (compare, e.g., to eqs.~(3.10) and~(2.7) of \citet{Wagoner91}, and eq. (9) of \citet{Friend83}).

The analyses of \citet{Wagoner91} and of \citet{karp77} are "local".
That is, it is assumed there that as the photon's frequency is shifted by an amount comparable to the strong line separation, the parameters of the plasma within which it propagate do not change \citep[the][assumptions are in fact more restrictive]{karp77}.
We have made the same assumption in deriving the final result of eq.~(\ref{eq:l_tau}).
However, the validity of this assumption is not obvious in our case. For the self-similar ejecta profiles described in \S~\ref{ssec:non-rel_profile}, $\rho$ and $T$ are steeply falling functions of $v_f$, $\rho$ is roughly proportional to $v_f^{-10}$ and $T$ is roughly proportional to $v_f^{-3}$.
Thus, as the photon moves outwards and its frequency is shifted (in the plasma frame) by $\sim v/c$ (i.e. by $dr/r\sim1$), $\rho$ and $T$ drop by factors of $10^3$ and $10^1$ respectively. This implies that $\tau_l$ may be significantly smaller than given by eq.~(\ref{eq:l}).

\section{Removing the effects of reddening}
\label{sec:extinction}

We show in this section that the effects of reddening on the observed UV/O signal may be removed using the UV/O light curves. This is particularly important for inferring $R_*$, since $R_*\propto T_{\rm col}^\alpha$ with $4\lesssim\alpha\lesssim5$ (see eqs.~\ref{eq:non_rel_T_ph},~\ref{eq:EffModel-THe} and~\ref{eq:EffModel-TCO}).

The model specific intensity, $f_\lambda$, is given by
\begin{equation}\label{eq:f_lambda}
    f_\lambda(\lambda,t)=\left(\frac{r_{\rm ph}}{D}\right)^{2}\sigma T_{\rm ph}^4 \frac{T_{\rm col}}{hc}
    g_{BB}(hc/\lambda T_{\rm col}) e^{-\tau_\lambda},
\end{equation}
where
\begin{equation}\label{eq:g_BB}
    g_{BB}(x)=\frac{15}{\pi^4} \frac{x^5}{e^x-1},
\end{equation}
$D$ is the distance to the source, and $\tau_\lambda$ is the extinction optical depth at $\lambda$.
Let us define $t_\lambda(t,\lambda)$ by
\begin{equation}\label{eq:t_lambda}
    \lambda T_{\rm col}[t=t_\lambda(t,\lambda)]=\lambda_0 T_{\rm col}(t),
\end{equation}
for some chosen $\lambda_0$. With this definition, the scaled light curves,
\begin{eqnarray}
\label{eq:rescaled_flux}
    \tilde{f}_{\lambda}[\lambda,t_\lambda(t,\lambda)]\equiv
    &&\left[\frac{D}{r_{\rm ph}(t_\lambda)}\right]^{2}
    \left[\frac{T_{\rm col}(t_\lambda)}{T_{\rm ph}(t_\lambda)}\right]^{4}
    \left[\frac{T_{0}}{T_{\rm col}(t_\lambda)}\right]^{5} \nonumber\\
    &\times& f_{\lambda}\left(\lambda,t_\lambda\right)
\end{eqnarray}
(where $T_0$ is an arbitrary constant) are predicted to be the same for any $\lambda$ up to a factor $e^{-\tau_\lambda}$,
\begin{equation}\label{eq:scaled_model}
    \tilde{f}_{\lambda}[\lambda,t_\lambda(t,\lambda)]=\sigma T_0^4 \frac{T_0}{hc}g_{BB}[hc/\lambda_0 T_{\rm col}(t)]\times e^{-\tau_\lambda}.
\end{equation}

Let us consider now how the scalings defined above allow one to determine the relative extinction in cases where the model parameters $\{E,M,R_*\}$ are unknown, and hence $\{T_{\rm col},T_{\rm ph},r_{\rm ph}\}(t)$, which define the scalings, are also unknown. For simplicity, let us first consider the case where the time dependence of the photospheric radius and temperature are well approximated by power-laws,
\be
\label{eq:power-laws}
    r_{\rm ph}\propto t^{\alpha_r}, \quad T_{\rm ph}\propto t^{-\alpha_T},
\ee
and the ratio $T_{\rm col}/T_{\rm ph}$ is independent of time. This is a good approximation for the time dependence of $r_{\rm ph}$ in general, and for the time dependence of $T_{\rm col}$ and $T_{\rm ph}$ for $T_{\rm ph}>1$~eV (see eqs.~\ref{eq:non_rel_r_ph},~\ref{eq:non_rel_T_ph},~\ref{eq:EffModel-THe},~\ref{eq:EffModel-r},~\ref{eq:EffModel-TCO}, and~\ref{eq:f_T}). In this case eq.~(\ref{eq:t_lambda}) gives
\begin{equation}\label{eq:t_scale}
    t_\lambda(t,\lambda)=\left(\frac{\lambda}{\lambda_0}\right)^{1/\alpha_T}t\,,
\end{equation}
and eq.~(\ref{eq:rescaled_flux}) may be written as
\begin{eqnarray}\label{eq:f_scale}
    \tilde{f}_{\lambda}[\lambda,t_\lambda(t,\lambda)]&=&{\rm Const.}\times
    \left(\frac{\lambda}{\lambda_0}\right)^{(-2\alpha_r+5\alpha_T)/\alpha_T} t^{-2\alpha_r+5\alpha_T}\nonumber\\
    &\times& f_{\lambda}\left[\lambda,\left(\frac{\lambda}{\lambda_0}\right)^{1/\alpha_T}t\right]\,.
\end{eqnarray}
The value of the constant that appears in eq.~(\ref{eq:f_scale}), for which the normalization of $\tilde{f}_{\lambda}$ is that given by eq.~(\ref{eq:scaled_model}), is not known, since it depends on the model parameters $\{E,M,R_*\}$. However, for any choice of the value of the constant, $\tilde{f}_{\lambda}$ defined by eq.~(\ref{eq:f_scale}) is predicted by the model to be given by eq.~(\ref{eq:scaled_model}) up to a wavelength independent multiplicative factor. Thus, the ratio of the scaled fluxes defined in eq.~(\ref{eq:f_scale}) determines the relative extinction,
\begin{equation}\label{eq:redenning}
    \frac{\tilde{f}_{\lambda}[\lambda_1,t_\lambda(t,\lambda_1)]}
    {\tilde{f}_{\lambda}[\lambda_2,t_\lambda(t,\lambda_2)]}= e^{\tau_{\lambda_2}-\tau_{\lambda_1}}\, .
\end{equation}

Let us consider next the case where the time dependence of $T_{\rm col}$ and $T_{\rm ph}$ is not a simple power-law. We have shown in \S~\ref{sec:const_opacity_model} and in \S~\ref{sec:real_model} that $T_{\rm col}$ and $T_{\rm ph}$ are determined by the composition and progenitor radius $R_*$, and are nearly independent of $E$ and $M$. Adopting some value of $R_*$, eq.~(\ref{eq:t_lambda}) may be solved for $t_\lambda(t,\lambda;R_*)$ and eq.~(\ref{eq:rescaled_flux}) may be written as
\begin{eqnarray}\label{eq:f_scale_p}
    \tilde{f}_{\lambda}[\lambda,t_\lambda(t,\lambda;R_*)]&=&{\rm Const.}\times t^{-2\alpha_r}
    T_{\rm ph}(t_\lambda)^{-4}T_{\rm col}(t_\lambda)^{-1} \nonumber\\
    &\times& f_{\lambda}\left(\lambda,t_\lambda\right).
\end{eqnarray}
The model predicts therefore that scaling the observed flux densities using the correct value of $R_*$, the observed light curves at all wavelengths should be given by eq.~(\ref{eq:scaled_model}), up to a multiplicative wavelength independent constant. For this value of $R_*$, the ratio of the scaled fluxes at different wavelengths is independent of $t$ and given by eq.~(\ref{eq:redenning}). The value of $R_*$ may be therefore determined by requiring the ratios of scaled fluxes to be time independent, and the relative extinction may then be inferred from eq.~(\ref{eq:redenning}). We use this method in \S~\ref{ssec:Applic_08D} for determining $R_*$ and the extinction curve for SN 2008D.

\section{Comparison to observations and simulations: RSG \& BSG progenitors}
\label{ssec:Applic_SG}

\subsection{SN1987A- A BSG progenitor}
\label{ssec:BSG}

\begin{figure}
\hspace{-5pt}
\includegraphics[scale=1.0]{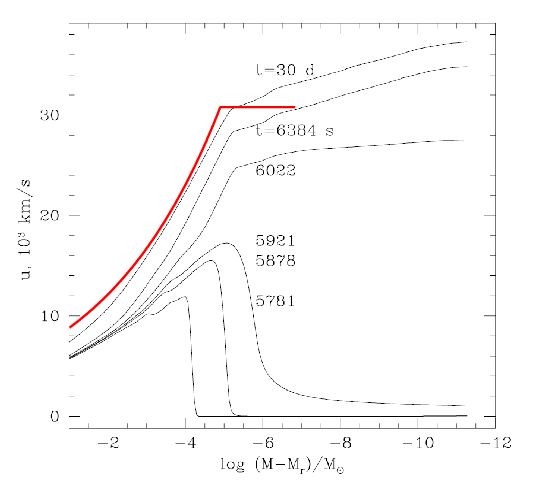}
\caption{
    Ejecta velocity profiles at different times, pre- and post- shock breakout, obtained in the 14E1.3 model calculation of \citet{Blinnikov00}, compared with the \citet{MM} approximation (red line) for the post-breakout velocity profile used in our analytical model, given by eq.~(\ref{eq:vs_nonrel}) for $\delta_m>\dmsb$ and assumed uniform at smaller $\delta_m$ (see eq.~\ref{eq:dmSB}).
    \label{fig:BlinnikovProf}
}
\end{figure}

Following the observations of SN 1987A, many numerical calculations modeling its light curve have been carried out
\citep[see e.g.][]{Hauschildt94}. The latest and most comprehensive of these calculations was carried out by \citet{Blinnikov00}, and it provides UBV light curves from the time of breakout to several months following the explosion. The \citet{Blinnikov00} radiation-hydrodynamics calculation, which includes a detailed treatment of the opacities and a multi-group transport approximation for the propagation of radiation, should, to our understanding, capture all the relevant physics.

\begin{figure}
\hspace{-5pt}
\includegraphics[scale=1.0]{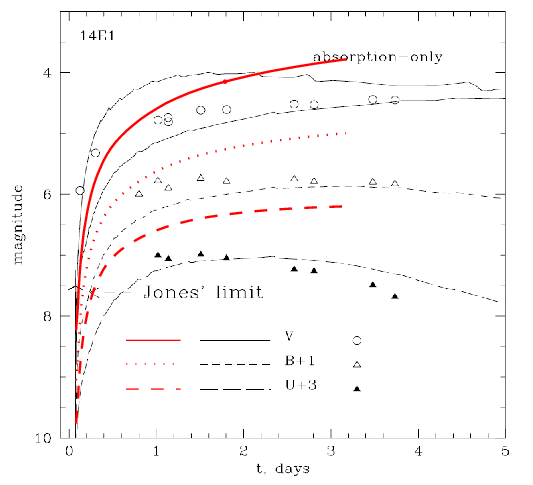}
\caption{Comparison of early UBV measurements of SN 1987A with the 14E1 model calculations of \citet{Blinnikov00} and with our model calculations (red line) for similar model parameters ($E_{B-V}=0.15$ and a distance modulus of 18.5 assumed). The uppermost black curve in the figure \citep[adapted from][]{Blinnikov00} is the V flux obtained in their 14E1A model calculation.
\label{fig:BlinnikovLC}
}
\end{figure}

In figures~\ref{fig:BlinnikovProf} and~\ref{fig:BlinnikovLC} we compare the results of our simple model to those of the detailed numerical calculations of \citet{Blinnikov00}. We use the same progenitor parameters as those used in \citet{Blinnikov00}: A BSG (H envelope with $n=3$, $f_\rho=1$) of radius $R_*=3.37\times 10^{12}\cm$, ejecta mass $M=14.67 M_{\odot}$, explosion energy $E=1.03, 1.34 \times 10^{51}\erg$ for models 14E1 and 14E1.3 respectively \citep[and composition as in the outer part of the progenitor given in figure 2 of][]{Blinnikov00}.
Figure~\ref{fig:BlinnikovProf} compares the numerical velocity profile with the \citet{MM} approximation we use in our model, given by eq.~(\ref{eq:vs_nonrel}). The two agree to better than 10\% over the relevant envelope mass fraction. In figure~\ref{fig:BlinnikovLC} we compare the numerical early UBV light-curves of \citet{Blinnikov00} with the ones calculated in our model, using eq.~(\ref{eq:tau_exact}) with the OP opacity \citep{OPCD}. As can be seen in the figure, our analytic model gives fluxes which are larger by a factor of $\sim 2$ than those of the numerical calculation. This difference may be due to differences between the OP opacities and those used by \citet{Blinnikov00}. The opacity given in \citet{Blinnikov98} for $\rho = 10^{-13}\, \DenUnit, T = 15000 \TempUnit $ and solar metalicity is larger than the OP opacity by roughly a factor of two. If a similar difference exists for the modified metalicity used in the SN1987A calculations, it would explain the luminosity discrepancy since the luminosity is roughly inversely proportional to $\kappa$, see eq.~(\ref{eq:L_BSG}).

\subsection{SNLS-04D2dc- A RSG progenitor}
\label{ssec:RSG}

\begin{figure}
\includegraphics[scale=1.0]{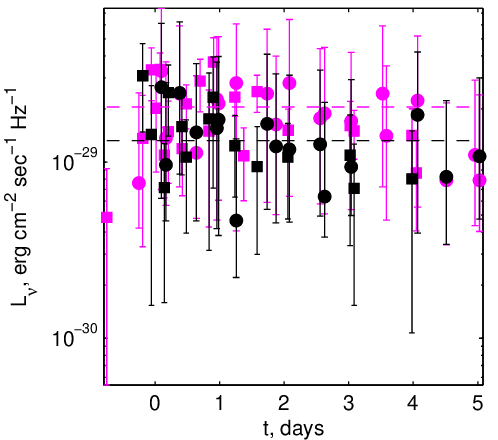}
\caption{
\label{fig:Obs-SNLS-04D2dc}
   Galex FUV ($\leff = 1539 \AA$ black) and NUV ($\leff = 2316 \AA$ magenta) observations of SNLS-04D2dc, not corrected for host and Galactic extinction.
   The photometric analyses of \citet{Gezari08} and of \citet{Schawinski08} are shown in circles and squares respectively.
   The dashed (black, magenta) lines show the (FUV, NUV) background levels inferred by \citet{Schawinski08}.
}
\end{figure}

The GALEX FUV and NUV measurements of the type IIp SNLS-04D2dc are shown in figure \ref{fig:Obs-SNLS-04D2dc}.
Given the relatively low signal to noise ratio, we do not attempt here to constrain the progenitor parameters using a detailed analysis of the UV emission (as we do for SN2008D in \S~\ref{ssec:Applic_08D}).
Rather, we show that the observed UV flux is consistent with that expected from an expanding shock heated envelope of a RSG progenitor, and compare our simple model predictions to those obtained using detailed numerical calculations.
For the latter purpose, we use the numerical calculations described in \citet{Gezari08}.

\citet{Gezari08} performed detailed numerical calculations, aimed at reproducing the early UV/O emission from the SNLS-04D2dc, the progenitor of which is most likely a RSG. Their calculation was performed in two stages. At the first stage, a hydrodynamic calculation of the explosion was performed using the one-temperature Lagrangian radiation hydrodynamics code KEPLER \citep{Weaver78}. At the second stage, the emission of radiation at time $t$ was calculated by solving, using the multi-group radiation transport code CMFGEN \citep{Dessart05}, the steady state radiation field for the hydrodynamic profiles obtained at time $t$, keeping the temperature profile fixed for $\tau>20$ and allowing it to self-consistently change at smaller optical depths.

\begin{figure}
\includegraphics[scale=1.0]{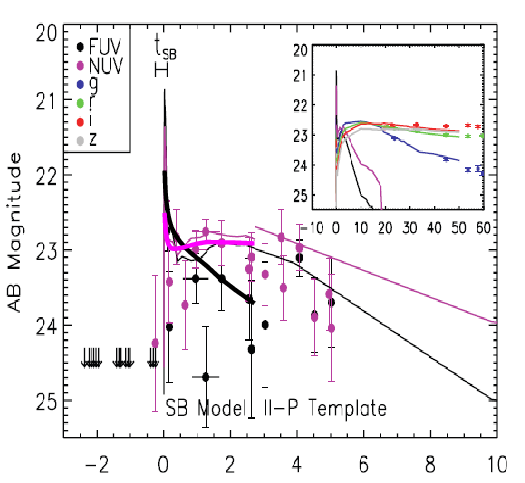}
\caption{
\label{Flo:GezariLC}
   Comparison of the numerical model calculations of \citet{Gezari08} (thin lines) with our model calculations (over laid thick lines) for similar progenitor parameters and extinction (see text for details). Due to the computationally demanding nature of the numerical calculation, the numerical model of \citet{Gezari08} extends up to the time marked by the vertical dotted line (the thin curves at later times are a scaled SN II-P template).
}
\end{figure}

In figure~\ref{Flo:GezariLC} we compare the \citet{Gezari08} calculations with the results obtained by our model, using eq.~(\ref{eq:tau_exact}) with the OP opacity \citep{OPCD}. The progenitor parameters we use are: A RSG (H envelope with $n=3/2$ and $f_\rho=25$) of radius $R_*=865 R_{\odot}$, explosion energy $E_{51}=1.44$ and ejecta mass of $M=8.9 M_{\odot}$. The stellar radius is similar to that used by \citet{Gezari08}, and $f_\rho=25$ was chosen (based on private communication) to provide an approximate description of the outer RSG envelope profile used in their calculation. The explosion energy and ejecta mass used in our model are 20\% larger and smaller respectively than those used by \citet{Gezari08}, i.e. the $E/M$ ratio is 40\% larger in our calculation.
This value was chosen to reproduce the observed luminosity. The light curves calculated by \citet{Gezari08} were shifted by $\sim-1.5$ magnitudes, i.e. the calculated luminosity was increased by a factor of $\sim 4$, to fit the observations. Since the luminosity is approximately proportional to $E/M$ (see eq.~\ref{eq:L_RSG}), this implies that for a given $E/M$ ratio our calculation predicts a luminosity that is larger than that of \citet{Gezari08} by a factor of $\sim3$ \citep[we have verified this by comparing the results of][to our model results for the same $E/M$ ratio]{Gezari08}.

The following point should be noted here.
Due to the large radius of the progenitor, the photosphere lies within the "breakout shell", i.e. $\delta_{m, ph.}<\dmsb$ (see eq.~\ref{eq:dmSB}), up to $t=t_{\rm BO}\simeq1.5$~d (see eq.~\ref{eq:t_bo}).
Our simple model is not valid at $t<t_{\rm BO}$.
However, we expect it to provide a reasonable approximation for the photospheric temperature and radius also at $t<t_{\rm BO}$, for the following reason.
As long as the photosphere lies at $\delta_{m, ph.}<\dmsb$, the diffusion time at the photosphere is short compared to $t$.
In this case we expect the spatial dependence of the radiation pressure to approximately follow $p\propto\tau$ (see \S~\ref{ssec:chck_adiabatic}), in which case the photospheric temperature is given by $aT_{\rm ph}^4=3p(\tau)/\tau$. Since, as explained in \S~\ref{ssec:chck_adiabatic}, the adiabatic pressure profile, $p\propto\tau^{1.1}$, is similar to that obtained for short diffusion time, $p\propto\tau$, and since the adiabatic pressure profile is valid at all times for $\delta_m\gg\dmsb$, we expect eq.~(\ref{eq:non_rel_T_ph}) to provide a good approximation for $T_{\rm ph}$ also at $t<t_{\rm BO}$. The temporal dependence of $r_{\rm ph}$ is expected to be somewhat stronger, at $t<t_{\rm BO}$, than the $r_{\rm ph}\propto t^{0.8}$ predicted by the simple adiabatic model, since the velocity profile at $\delta_{m, ph.}<\dmsb$ is shallower than predicted by eq.~(\ref{eq:vs_nonrel}).
The largest deviation from the $r_{\rm ph}\propto t^{0.8}$ behavior would be obtained assuming uniform velocity at $\delta_{m, ph.}<\dmsb$, which would yield $r_{\rm ph}\propto t$.

The factor of $\sim2$ discrepancy between the luminosity predicted by our model and that obtained by \citet{Gezari08} is due to the fact that our model predicts a somewhat, $\sim40$\%, larger velocity for the fast outer shells, and hence a larger photospheric radius. In a subsequent publication \citep{RabinakIP} we examine the accuracy of the approximate ejecta density and velocity profiles described in \S~\ref{ssec:non-rel_profile} for a wide range of progenitor models.

\section{The early UV/O emission of SN 2008D}
\label{ssec:Applic_08D}

The analysis of the early UV/O emission of SN2008D is complicated by two major factors. First, the extinction is large. It was loosely constrained by \citet{Alicia08D} to $0.4<E(B-V)<0.8$. This large extinction makes it difficult to extract the color temperature from the observations, and increases the uncertainty of the interpretation. The second complication arises from the fact that there is only one set of measurements in the UV at $t\lesssim 2$~days, and most of the relevant data points are at $t>1$~d. This implies, as explained in \S~\ref{ssec:var_opacity}, that the simple approximate solutions given by eqs.~(\ref{eq:EffModel-r}) and~(\ref{eq:EffModel-TCO}) for the photospheric radius and temperature are not accurate. For both He and mixed He-C/O compositions the reduction in opacity due to He recombination at $\gtrsim1$~d implies that the photosphere penetrates into the envelope beyond the range of validity of the approximation of eq.~(\ref{eq:init_dnsty_prfl}) for the initial density profile. For the analysis of SN 2008D, we will use therefore a more detailed description of the density and pressure profiles of the ejecta. We first describe the modified ejecta profiles we use in \S~\ref{ssec:mod_ejecta}, and then analyze the SN2008D observations using our model in \S~\ref{ssec:08_const}. As we show there, the deviation of the ejecta profiles from the $\delta_m\rightarrow0$ self-similar profiles described in \S~\ref{ssec:non-rel_profile} lead to modifications of the model predictions at $t\gtrsim2$~d. Comparison of our results to some earlier work appears in \S~\ref{ssec:comp}.

\subsection{Modified ejecta profiles}
\label{ssec:mod_ejecta}

\begin{figure}
\hspace{-5pt}
\includegraphics[scale=1.0]{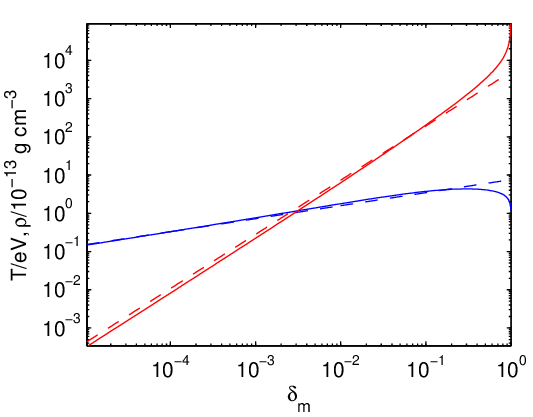}
\caption{
\label{fig:TempDmA}
    A comparison of the temperature and density profiles (in blue and red respectively) of the ejecta given by the self-similar solution of  \S~\ref{sec:const_opacity_model} (dashed lines) with those obtained in the "Harmonic-mean" model described in \S~\ref{ssec:mod_ejecta} (solid lines), for an explosion of an $n=3$ envelope with $E_{51} = 6$, $M= 7 M_{\odot}$ and $R_* = 10^{11}$~cm, at $t=10^5\sec$.
}
\end{figure}

The derivations of $r_{\text{ph}}$ and $T_{\text{ph}}$ in \S~\ref{ssec:prop_photo} and in \S~\ref{sec:real_model} are based on  the self-similar model of \S~\ref{ssec:non-rel_profile} for the density and pressure profiles of the ejecta. This model, in which $p(v_f)$ and $\rho(v_f)$ are both roughly proportional to $v_f^{-10}$, is valid in the limit $\delta_m\rightarrow0$ ($v_f\rightarrow\infty$). In order to extend the model to larger values of $\delta_m$, we adopt the "Harmonic-Mean" model suggested by \citet{MM}. In this model, the density and pressure profiles are obtained by interpolating between the small $\delta_m$ (large $v_f$) self-similar power-law dependence of $p(v_f)$ and $\rho(v_f)$, and the power-law dependence $p(v_f)\propto v_f^\alpha$, $\rho(v_f)\propto v_f^\beta$ with $\alpha\approx2$ and $\beta\approx-1$, obtained in the approximate analysis of \citet{Chevalier89} for the lower velocity ejecta. This power-law dependence is obtained by assuming that the shock propagating within the ejecta may be approximately described, for large $\delta_m$, by the Primakoff self-similar solution \citep[a particular, analytic, case of the Sedov-von Neumann-Taylor solutions for shock propagation into $\rho\propto r^{-\omega}$ density profiles, obtained for $\omega=17/7$, e.g.][]{Gaffet84,Bernstein80}, and by an approximate (self-similar) description of the post-breakout acceleration of the shocked plasma. The two power-law solutions describing the large and small $v_f$ behavior of the density are matched in the \citet{MM} "Harmonic-Mean" model at $\rho=\rhob$ and $v_f=\vrhob$, and the pressure profiles at $p=\pb$ and $v_f=\vpb$ \citep[see eqs.~(46) and ~(47) of][]{MM}. $\rhob$ and $\vrhob$ are determined by requiring the ejecta mass and (kinetic) energy to equal $M$ and $E$ respectively. $\pb$ and $\vpb$ are determined in \citep{MM} by examining numerical simulation results. We find that their parameter choice of $\pb$ and $\vpb$ leads an to over-estimate of the temperature, compared to that of the self-similar $\delta_m\rightarrow0$ solution, by $\sim15\%$ at $\delta_m=10^{-3}$. We therefore modify the value of $\pb$ to obtain the correct self-similar behavior at $\delta_m\rightarrow0$.

The "Harmonic-mean" density and temperature profiles obtained as described in the preceding paragraph are compared in fig.~\ref{fig:TempDmA} with those of the $\delta_m\rightarrow0$ self-similar solution. Significant deviation from the self-similar profiles is obtained for $\delta_m \gtrsim 0.1$. As discussed in \S~\ref{ssec:08_const}, this deviation affects the model predictions for $t\gtrsim2$~d. The accuracy of the "Harmonic-mean" model was examined in \citep{MM} by comparing it to the results of numerical calculations of the explosions of various RSG and BSG progenitors. Since it was found that this analytic model provides a good approximation for the envelope's profiles for different initial density structures of the progenitors, we expect the Harmonic-mean model to provide a good approximation also for the SN2008D envelope profiles. However, additional work, which is beyond the scope of the current paper, is required in order to obtain a quantitative estimate of the accuracy of the approximation for $\delta_m>0.1$.

\subsection{Models vs. observations}
\label{ssec:08_const}

Spectroscopic observations have constrained the fraction of Hydrogen in the ejecta to $\lesssim5\times10^{-4}M_{\odot}$ \citep{Tanaka09}. We therefore consider below He and He-C/O envelopes. As explained in \S~\ref{sec:extinction}, the relative extinction may be inferred from the light curves at different frequencies. However, for the clarity of the presentation we first analyze the data using two relative extinction curves that differ significantly in their short wavelength behavior, a Milky Way extinction curve with $R_v =3.1$ (here after MW) and a Small Magellanic Cloud extinction curve (hereafter SMC) \citep{cardelli}, and only later show how the extinction curve may be directly inferred from the data.

\begin{figure*}
\includegraphics[scale=1]{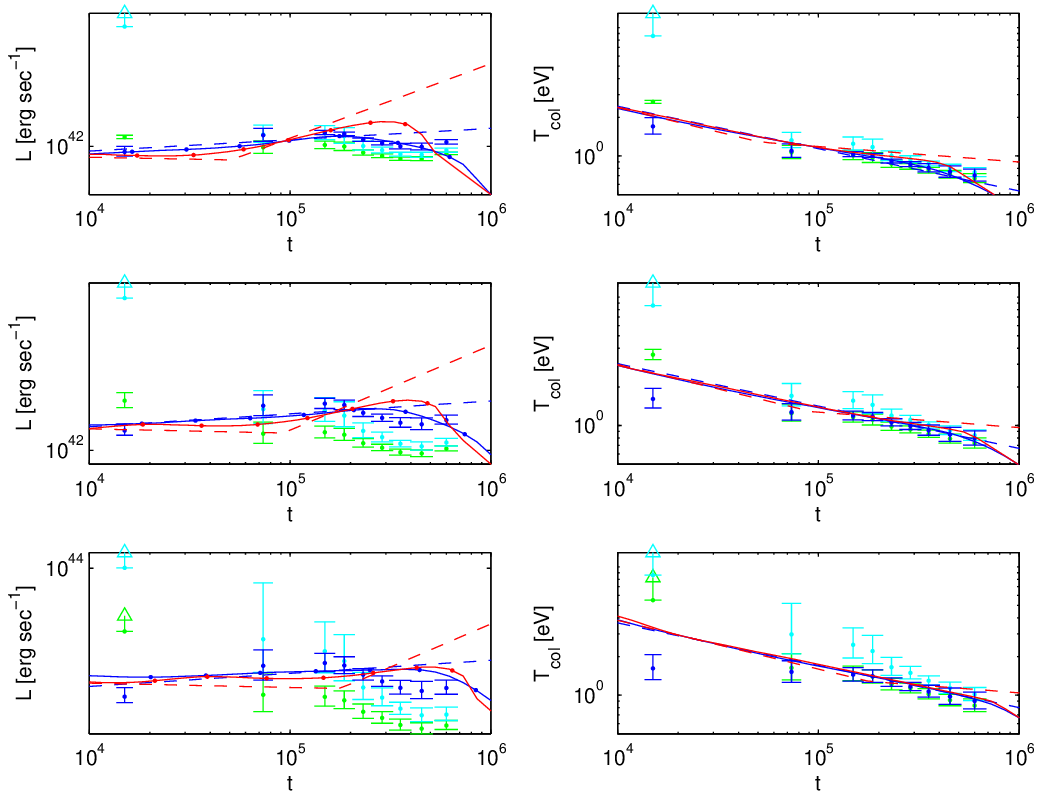}
\caption{A comparison of the color temperature $T_{\rm col}$ and bolometric luminosity $L$ inferred from the data with those obtained in our model for different progenitor and explosion parameters. Model results are shown for three progenitor radii, $R_*={1,3,10}\times10^{11}$~cm in the top, middle, and bottom panels respectively, and for two envelope compositions, He dominated (red) and mixed He-C/O composition with a He mass fraction $1-Z=0.3$ (blue). $E_{51}=6$ and $M/M_\odot=7$ (and $D=27$~Mpc) are assumed for all models. Shown are both the simple analytic approximations for $L$ and $T_{\text{col}}$ given in \S~\ref{ssec:var_opacity}, which are based on the self-similar ejecta profiles of \S~\ref{ssec:non-rel_profile} (dashed lines), and $L$ and $T_{\text{col}}$ obtained solving eqs.~(\ref{eq:tau_exact}),~(\ref{eq:r_thrm}) and~(\ref{eq:r_diff}) with the OP opacity tables for the modified ejecta profiles described in \S~\ref{ssec:mod_ejecta} (full lines). The observed, extinction corrected $L(t)$ and $T_{\text{col}}(t)$ are inferred from the data by assuming a specific extinction curve (MW, green data points, or SMC, cyan data points), and determining the absolute value of the extinction by requiring $T_{\text{col}}$ inferred from the observations to agree with model prediction at $t\sim2{\,\rm d}$. $E(B-V)=0.625, 0.7, 0.8$ are inferred for $R_*={1,3,10}\times10^{11}$~cm respectively. Blue data points are obtained for (best fit) extinction curves which are determined from the data itself, for the mixed He-C/O composition (see fig.~\ref{fig:SN08DExtR1E6M7}).
\label{fig:08D}
}
\end{figure*}

Figure~(\ref{fig:08D}) presents a comparison of the color temperature $T_{\rm col}$ and bolometric luminosity $L$ inferred from the data with those obtained in our model for different progenitor and explosion parameters. We use the observations of \emph{Swift}/UVTO (V, B, U, UVW1, UVM2, UVW2), Palomar (g, r, i, z) \citep[both taken from][]{Alicia08D}, and FLWO (B, V, r, i) \citep{Modjaz09}. Since observations by different telescopes were carried out at different times, we interpolate the observations in different bands to times close to the observation times of the \emph{Swift}/UVTO and the Palomar telescopes. In order to derive $T_{\text{col}}(t)$ and $L(t)$ from the observations, the measured fluxes should be corrected for extinction. We carry out this correction by (i) assuming a specific extinction curve (MW or SMC) and (ii) determining the absolute value of the extinction by requiring $T_{\text{col}}$ inferred from the observations to agree with the model prediction at $t\sim2{\,\rm d}$. This implies that for each set of model parameters ($E$, $M$, $R_*$, envelope composition), a different value of the absolute extinction, $E(B-V)$ is chosen.

For all the models shown in the figure we have used $E_{51}=6$ and $M/M_\odot=7$, as suggested by \citet{Mazzali08} from the spectral analysis of the observations at maximum light. As shown in \S~\ref{sec:const_opacity_model} and in \S~\ref{sec:real_model}, $E$ and $M$ (mainly the ratio $E/M$) determine the normalization of the model luminosity, see eqs.~(\ref{eq:L_He}) and~(\ref{eq:L_CO}), but do not affect the time dependence of the luminosity, and $T_{\text{col}}$ is nearly independent of $E$ and $M$, see eqs.~(\ref{eq:EffModel-THe}) and~(\ref{eq:EffModel-TCO}). Since the conclusions drawn from the comparisons in fig.~(\ref{fig:08D}) are based on the time dependence of $T_{\text{col}}$ and $L$, their validity is independent of the exact values chosen for $E$ and $M$.

\begin{figure*}
\includegraphics[scale=1.0]{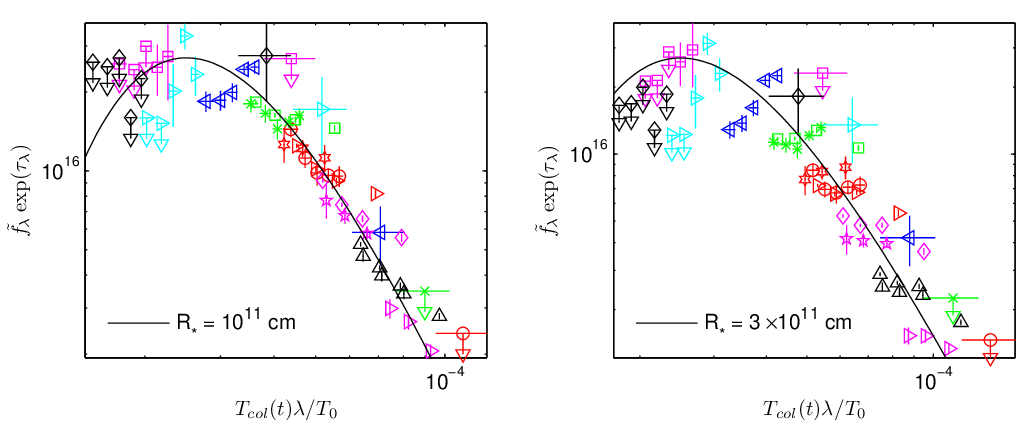}
\caption{The measured flux densities $f_\lambda$, corrected for extinction and scaled according to eq.~(\ref{eq:rescaled_flux}) (using $D=27$~Mpc and $T_0=1$~eV), compared with the model prediction of eq.~(\ref{eq:scaled_model}) (solid line) for the mixed He-C/O composition models presented in fig.~\ref{fig:08D} with $R_*=10^{11}$~cm and $R_*=3\times10^{11}$~cm. Different symbols and colors show measurements at different bands: red circle-- V, green x-- B, blue left triangle-- U, cyan right triangle-- UVW1, magenta square-- UVM1, black diamond-- UVW2, magenta pentagon-- r, red hexagon-- g, black triangle-- i, magenta right triangle-- z,
    \citep[taken from][]{Alicia08D};
    black triangle-- i', magenta diamond-- r', red right triangle-– B, green square-- V \citep[taken from][]{Modjaz09}. \label{fig:BBLC}
}
\end{figure*}

Fig.~(\ref{fig:08D}) shows $T_{\text{col}}$ and $L$ for models with three progenitor radii, $R_*={1,3,10}\times10^{11}$~cm, and two compositions, He dominated and mixed He-C/O composition with a He mass fraction $1-Z=0.3$. We show both the simple analytic approximations for $L$ and $T_{\text{col}}$ given in \S~\ref{ssec:var_opacity} (dashed curves), which are based on the self-similar ejecta profiles of \S~\ref{ssec:non-rel_profile}, and $L$ and $T_{\text{col}}$ obtained solving eqs.~(\ref{eq:tau_exact}),~(\ref{eq:r_thrm}) and~(\ref{eq:r_diff}) with the OP opacity tables for the modified ejecta profiles described in \S~\ref{ssec:mod_ejecta}. Since models with larger initial radii predict higher $T_{\text{col}}$, the absolute extinction inferred for models with larger radii is larger, $E(B-V)=0.625, 0.7, 0.8$ for $R_*={1,3,10}\times10^{11}$~cm respectively. Once the absolute extinction is determined, from the comparison of observed and predicted $T_{\text{col}}$ at $t\sim2{\,\rm d}$, $T_{\text{col}}(t)$ and $L(t)$ are inferred form the observations  using the two relative extinction curves (MW, SMC). In determining the observed $T_{\text{col}}$ and $L$, the photometric measurements are converted into monochromatic fluxes at the effective wavelength of the broad-band filters, and a BB temperature is determined by a least-square fit to these fluxes. The resulting $T_{\text{col}}$ and $L$ are shown in fig.~\ref{fig:08D}. The error bars represent the uncertainties obtained in the least-square fits.

Examining fig.~\ref{fig:08D} we infer a small progenitor radius, $R_*\approx10^{11}$~cm. Progenitors with larger radii require larger extinction to account for the observed flux distribution at $t=2$~day, which in turn imply that the extinction corrected $L(t)$ decreases with time at $t<2$~d, in contrast with the roughly time independent $L$ predicted by the models (see also eqs.~(\ref{eq:L_He}) and~(\ref{eq:L_CO})), and that the extinction corrected $T_{\rm col}$ decreases faster than predicted by the models for $t<2$~d. This is due to the fact that at earlier times the flux is dominated by shorter wavelength bands, for which the extinction correction is larger. Comparing model predictions and observations at $t>2$~d, we find that a mixed He-C/O composition is preferred over a He dominated one. The Z=0.7 model presented provides a good fit to the data. We find that Z$\sim$ few 10's of percent is required to fit the observations. Note, however, that the light curve at $t>2$~d depends on the non self-similar part of the density and pressure profiles of the ejecta, for which we have used the "Harmonic-mean" approximation described in \S~\ref{ssec:mod_ejecta}. As mentioned in \S~\ref{ssec:mod_ejecta}, additional work, which is beyond the scope of the current paper, is required in order to obtain a quantitative estimate of the accuracy of this approximation. We can not rule out, therefore, the possibility that the observations may be explained with a He dominated contribution and a density profile at large $\delta_m$, that deviated from that given by the "Harmonic-mean" approximation.

\begin{figure}
\includegraphics[scale=1]{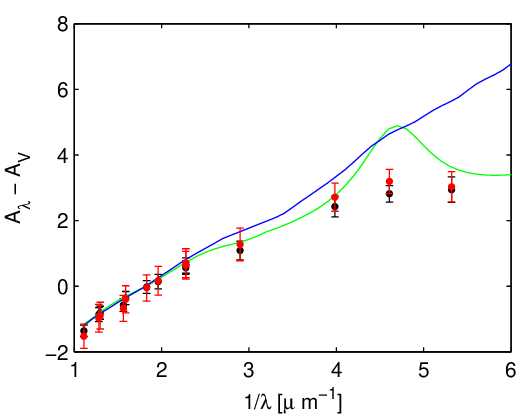}
\caption{
\label{fig:SN08DExtR1E6M7}
    Comparison of the extinction curves inferred from the models described in fig~\ref{fig:BBLC} (black and red points for $R_*=10^{11}$~cm and $R_*=3\times10^{11}$~cm respectively) with $A_V = 2.21$ MW and $A_V =2.16$ SMC extinction curves \citep{cardelli}. We find $E(B-V) = 0.63, 0.70$ and  $A_V = 2.39, 3.03$ for $R_*=10^{11}$~cm and $R_*=3\times10^{11}$~cm respectively.}
\end{figure}

As explained in \S~\ref{sec:extinction}, $R_*$ and the relative extinction between different wavelengths may be determined from the O/UV light curves by requiring that the light curves observed at different wavelengths should all be given, after scaling according to eqs.~(\ref{eq:t_lambda}) and ~(\ref{eq:f_scale}), by a single function, given by eq.~(\ref{eq:scaled_model}). In fig.~\ref{fig:BBLC} we compare the measured specific intensities, corrected for extinction and scaled according to eqs.~(\ref{eq:t_lambda}) and ~(\ref{eq:f_scale}), with the model prediction, eq.~(\ref{eq:scaled_model}). For the scaling we have used $\{r_{\rm ph}(t),T_{\rm col}(t),T_{\rm ph}(t)\}$ obtained for the mixed He-C/O composition models presented in fig.~\ref{fig:08D} with $R_*=10^{11}$~cm and $R_*=3\times10^{11}$~cm. The extinction $\tau_\lambda$ was obtained by requiring the scaled intensity to best fit that predicted by eq.~(\ref{eq:scaled_model}) (taking into account all data points at $t<4$~d). The resulting extinction curves are shown in fig.~(\ref{fig:SN08DExtR1E6M7}). As can been seen in the figure, the $R_*=10^{11}$~cm model provides a much better description of the data than the $R_*=3\times10^{11}$~cm model. The extinction curve is more compatible with a MW extinction than with SMC extinction. It differs from the MW curve at short wavelengths, showing no prominent graphite bump. This, as well as the values obtained for $A_V$ and $E(B-V)$, $A_V = 2.39$ and $E(B-V) = 0.63$, are consistent with the extinction inferred in \citep[][and private communication]{Alicia08D}.

The following point should be explained here. We have obtained the absolute values of the extinction by adopting some values for $E$ and $M$. It is important to note, that, as explained in detail in \S~\ref{sec:extinction}, the relative extinction is independent of $E$ and $M$, and may be inferred without making assumptions regarding their values. The model predicted temperature, $T_{\rm ph}$, and $T_{\rm col}/T_{\rm ph}$ are almost independent of $E$ and $M$, and depend only on $R_*$ (and on the composition). $E$ and $M$ determine the normalization of $r_{\rm ph}$, $r_{\rm ph}\propto E^{0.4}/M^{0.3}$ (see eq.~(\ref{eq:EffModel-r})), but do not affect its time dependence. Thus, modifying $E$ and $M$ changes the scaled fluxes of eq.~(\ref{eq:rescaled_flux}) by some multiplicative factor, which is wavelength independent. Thus, the ratios of the scaled fluxes at different wavelengths are independent of $E$ and $M$, and so are the inferred relative extinctions. Using the relative extinction inferred from the data, and adopting some relation between the relative and absolute extinctions, which determines the absolution extinction ($\tau_\lambda$) we may therefore constrain the $E^{0.4}/M^{0.3}$ ratio by comparing the predicted and measured (absolute) flux at some frequency. Using the inferred $E(B-V) = 0.63$ implies $A_V = 2.39$ for a MW extinction, for which we infer $E_{51}/(M/M_{\odot})=0.75$.

\begin{figure}
\includegraphics[scale=1.0]{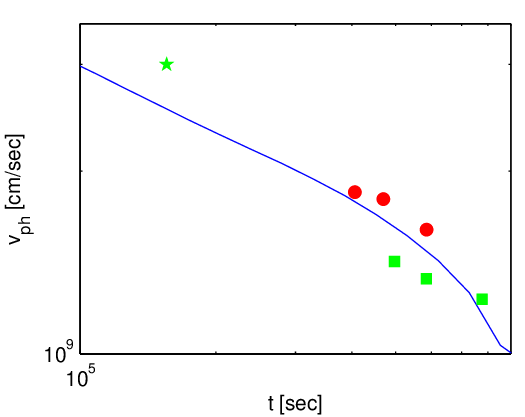}
\caption{
\label{fig:SN08DVelocityComp2Obs}
    Comparison of the photospheric velocity of the best fit model, in blue solid line, to results of other authors. In green squares are velocity interpreted from the He I $\lambda5876$ line, and in green pentagon is the velocity from the analysis done via SYNOW to the "W" feature which later on vanishes both taken from \citet{Modjaz09}. In red circles the photospheric velocity measured from spectral modeling by \cite{Tanaka09}.
}
\end{figure}

\subsection{Comparison with previous work}
\label{ssec:comp}

Based on our analysis of the early UV/O emission of SN2008D we infer a small progenitor radius, $R_*\approx10^{11}$~cm, an $E_{51}/(M/M_{\odot})\approx0.8$, a preference for a mixed He-C/O composition (with C/O mass fraction of 10's of percent), $E(B-V) = 0.63$ and an extinction curve given by fig.~\ref{fig:SN08DExtR1E6M7}. We compare below our conclusions to those of earlier analyses.

Explosion models for SN2008D were consider by \citet{Mazzali08} and by \citet{Tanaka09}. For $E$ and $M$ \citet{Mazzali08} infer $E_{51}\sim6$ and $M\sim7M_\odot$, i.e. $E_{51}/(M/M_{\odot})\sim0.85$ while \citet{Tanaka09} infer $E_{51}=6\pm 2.5$ and $M =5.3\pm 1 M_{\odot}$, and $E_{51}/(M/M_{\odot})$ in the range
$ 0.8 < E_{51}/(M/M_{\odot}) < 1.3$. These values are consistent with our inferred value of $E_{51}/(M/M_{\odot})\approx0.8$. The progenitor radius inferred from the \citep{Tanaka09} analysis is $0.9 \lesssim R_{*}/10^{11}{\rm cm} \lesssim 1.5 $, also consistent with our inferred  $R_*\approx10^{11}$~cm.

In fig.~\ref{fig:SN08DVelocityComp2Obs} we compare the photospheric velocity obtained in our model with those obtained by \citet{Modjaz09} and by \citet{Tanaka09} analyzing SN2008D spectra. As can bee seen in the figure, our model predictions are in good agreement with the results inferred from the spectral analyses. We also note that \citet{Mazzali08} infer, based on spectral analysis, that the mass of the ejecta shell moving at $> 0.1c$ is $\sim 0.03 M_{\odot}$, consistent with our model prediction of $\sim 0.02 M_{\odot}$ at this velocity.

Let us discuss next our conclusion that the He envelope contains a significant (10's of \%) C/O fraction.
This conclusion is consistent with the results of Mazzali et al., who find a large fraction, $\sim10$\% of C at the fast $v>25000{\rm km/s}$ He shells, rising to $\sim30$\% at the $v\simeq20000{\rm km/s}$ shells (P. Mazzali, private comm.). Both Mazzali et al. and \citet{Tanaka09} find a low, $\sim0.01$ mass fraction of O at the fast shells.

Finally, we comment on the analysis of \citet{Chev08}, who find a large progenitor radius, $R_*\approx10^{12}$~cm, based on both the X-ray and the UV/O emission.
\citet{Chev08} obtain $R_*\approx10^{12}$~cm by interpreting the X-ray emission as thermal, $L_X=4\pi R_*^2 \sigma T_X^4$, and adopting $T_X=0.36$~keV. Their motivation for a thermal interpretation of the X-ray emission, despite the fact that the spectrum is non thermal and extends beyond 10~keV, was that the non-thermal spectrum is difficult to explain theoretically.
As explained in the introduction, the non-thermal spectrum is a natural consequence of the physics of fast radiation mediated shocks \citep{Katz10}.
Thus, there is no motivation, and it is inappropriate, to infer $R_*$ assuming thermal X-ray emission. Moreover, it should be kept in mind that the X-ray breakout may take place within the wind surrounding the progenitor, at a radius significantly larger than $R_*$ \citep[e.g.][]{Waxman07}.
The discrepancy between their large radius and our small radius inferred from the UV/O data is due to the fact that we take into account the modification of the opacity due to He recombination and the difference between color and effective temperatures. Since $R_*\propto\kappa$, neglecting the reduction of opacity due to HE recombination at $T=1$~eV ($t\sim1$~d) leads to a significant overestimate of the radius (compare eqs.~(\ref{eq:R_H}) and~(\ref{eq:R_He})), and neglecting the difference between color and effective temperatures leads to an additional over estimate of the radius,
$R_*\propto T_{\rm col}^4=(T_{\rm col}/T_{\rm
ph})^4T_{\rm ph}^4$ with $(T_{\rm col}/T_{\rm ph})^4\approx 2$ (see eq.~(\ref{eq:R_He})).

\section{Conclusion}
\label{sec:Conclusion}

We have presented a simple model for the early UV/O emission of core collapse supernovae. The photospheric radius, $r_{\rm ph}(t)$, and (effective) temperature, $T_{\rm ph}(t)$ are given for H envelopes by eqs.~(\ref{eq:non_rel_r_ph}) and~(\ref{eq:non_rel_T_ph}) in \S~\ref{sec:const_opacity_model}, and for He and mixed He-C/O envelopes (including pure C/O envelopes) by eqs.~(\ref{eq:EffModel-THe})--(\ref{eq:EffModel-TransT}) in \S~\ref{ssec:var_opacity}. $T_{\rm ph}$ is determined by the composition and by the progenitor radius, $R_*$, and is nearly independent of the ejecta mass, $M$, and energy, $E$. $M$ and $E$ determine the normalization of $r_{\rm ph}(t)$, but not its time dependence, $r_{\rm ph}\propto E^{0.4}/M^{0.3}$. The bolometric luminosity predicted by the model is nearly time independent (for $T_{\rm ph}>1$~eV), see eqs.~(\ref{eq:L_RSG}),~(\ref{eq:L_BSG}),~(\ref{eq:L_He}) and ~(\ref{eq:L_CO}). Both $r_{\rm ph}(t)$ and $T_{\rm ph}(t)$ are only weakly dependent on $n$, the exponent determining the dependence of the progenitor's density on the distance from the edge of the star, see eq.~(\ref{eq:init_dnsty_prfl}). A discussion of the deviation of the
spectrum from a black-body spectrum is given in \S~\ref{ssec:calc_T_col}, where we find that the ratio of color to effective temperature is approximately constant at early time, $T_{\rm col}/T_{\rm ph}\approx1.2$ (see figs.~\ref{fig:EffModelHNew}--\ref{fig:EffModelHeCO-WR70New}).

For progenitor radii $R_*\lesssim10^{12}$~cm, $T_{\rm ph}$ approaches 1~eV on day time scale. For He envelopes, significant recombination takes place at $\sim1$~eV, leading to a significant reduction of the opacity, which, in turn, prevents $T_{\rm ph}$ from dropping significantly below 1~eV, since the photosphere penetrates (deeper) into the envelope up to the point where the temperature is high enough to sustain significant ionization (see fig.~\ref{fig:EffModelHe-WRNew}). Significant amounts of C/O in the envelope allow $T_{\rm ph}$ to drop below $\sim1$~eV, since these atoms are partially ionized at lower temperatures as well (see fig.~\ref{fig:EffModelHeCO-WR70New}). Model predictions depend only weakly on the C:O ratio.

Since $T_{\rm col}(t)$ is determined by the composition and by $R_*$, the progenitor radius may be inferred from the observed $T_{\rm col}$. Eqs.~(\ref{eq:R_H})--(\ref{eq:R_He-CO}) give $R_*$ as function of the observed $T_{\rm col}$ for H, He, and He-C/O envelopes. A few comments should be made at this point. For a space and time independent opacity, $R_*\propto \kappa T_{\rm col}^4$ (see eq.~(\ref{eq:non_rel_r_ph})). In case $\kappa$ varies with temperature and density, the appropriate value of $\kappa$, i.e. its value at the photosphere at the time $T_{\rm col}$ is measured, should be used. The fractional uncertainty in $R_*$ is similar to the fractional uncertainty in $\kappa$. The strong dependence of $R_*$ on $T_{\rm col}$ implies that relatively small uncertainties in the determination of $T_{\rm col}$ from the observations, or in its calculation in the model, lead to large uncertainties in the inferred $R_*$. Our approximate treatment of the deviation from black-body spectrum due to photon diffusion implies $T_{\rm col}$ is larger than $T_{\rm ph}$ by $\approx20\%$. Estimating the uncertainty in the magnitude of this effect to be comparable to the magnitude of the effect, implies a factor of $\approx2$ uncertainty in the inferred $R_*$. A more accurate estimate would require a more detailed treatment of photon transport (including the effects of effective line opacity enhancement due to the large velocity gradients, see \S~\ref{ssec:exp_opacity}).

Uncertainties in the observational determination of $T_{\rm col}$ are due to reddening. As explained in detail in \S~\ref{sec:extinction}, $R_*$ and the relative extinction at different wavelengths may be inferred from the UV/O light curves. Scaling the observed light curves at different frequencies according to eq.~(\ref{eq:f_scale_p}), with $T_{\rm col}(t)$ and $T_{\rm ph}(t)$ obtained in a model with the correct value of $R_*$, should bring all the light curves to coincide, up to a factor $e^{-\tau_\lambda}$ where $\tau_\lambda$ is the extinction optical depth. The value of $R_*$ may therefore be determined by requiring the ratios of scaled fluxes to be time independent, and the relative extinction between two wavelengths may then be inferred from value of this ratio (see eq.(\ref{eq:redenning})). For the case where the time dependence of the photospheric radius and of the photospheric and color temperatures are well approximated by power-laws, which is a good approximation for the time dependence of $r_{\rm ph}$ in general and for the time dependence of $T_{\rm col}$ and $T_{\rm ph}$ for $T_{\rm ph}>1$~eV, the relative extinction may be inferred independently of $R_*$, from the ratio of the fluxes scaled according to eq.~(\ref{eq:f_scale}).

In \S~\ref{ssec:Applic_SG} we have compared our model predictions to observations of the early UV/O emission available for two SNe (SN1987A, SNLS-04D2dc), arising from RSG and BSG progenitors, and to detailed numerical (radiation transport-hydrodynamics) simulations, that were constructed to reproduce these observations \citep{Blinnikov00,Gezari08}. We have shown that our simple model may explain the observations. We find, however, that our predicted luminosity is $\approx2$ times larger than obtained (for similar progenitor and explosion parameters) by the detailed numerical simulations. In the case of the BSG SN, this discrepancy is probably due to differences in the opacity tables we use and those used in the simulations ($L\propto1/\kappa$, see \S~\ref{ssec:BSG}). In the case of the RSG SN, the discrepancy is due to the fact that our model predicts a somewhat ($\sim40$\%) larger velocity for the fast outer shells, and hence a larger photospheric radius than that obtained in the numerical simulation. Since the details of the explosion model of \citet{Gezari08} are not given in their paper, it is difficult to determine the source of this discrepancy. In a subsequent publication \citep{RabinakIP}, we will examine the accuracy of the approximate ejecta density and velocity profiles described in \S~\ref{ssec:non-rel_profile} for a wide range of progenitor models.

In \S~\ref{ssec:Applic_08D} we have used our model to analyze the early UV/O measurements of SN2008D. For this explosion we infer a small progenitor radius, $R_*\approx10^{11}$~cm, an $E_{51}/(M/M_{\odot})\approx0.8$, a preference for a mixed He-C/O composition (with C/O mass fraction of 10's of percent), $E(B-V) = 0.63$ and an extinction curve given by fig.~\ref{fig:SN08DExtR1E6M7}.
Our results are consistent (see discussion in \S~\ref{ssec:comp}) with the $E/M$ values inferred from modeling the light curve and spectra at maximum light \citep{Mazzali08,Tanaka09}, with the $R_*$ range obtained in the stellar evolution models described in \citet{Tanaka09}, and with the extinction inferred from the analyses of spectra at maximum light \citep[e.g.][]{Alicia08D}. The photospheric velocity predicted by our model is consistent with the velocities inferred from spectral analyses \citep{Mazzali08,Modjaz09,Tanaka09}, see fig.~{\ref{fig:SN08DVelocityComp2Obs}.

Our conclusion that the He envelope contains a significant (10's of \%) C/O fraction is not as robust as the other conclusions, since it relies on observations at $t>2$~d, where the emission is dominated by shells that were initially located at distances from the edge of the star for which the asymptotic (self-similar) description of the density, eq.~(\ref{eq:init_dnsty_prfl}), does not hold (see detailed discussion in \S~\ref{ssec:08_const}). In order to describe the light curve at $t>2$~d we have used the approximation described in \S~\ref{ssec:mod_ejecta} for the density and pressure profiles of the ejecta. Additional work, which is beyond the scope of the current paper, is required in order to obtain a quantitative estimate of the accuracy of this approximation. We can not rule out, therefore, the possibility that the observations may be explained with a He dominated composition and a density profile that deviates from the approximation used. Nevertheless, our conclusion is consistent with the analysis of Mazzali et al., who find a large fraction, $\sim10$\% of C at the fast $v>25000{\rm km/s}$ He shells, rising to $\sim30$\% at the $v\simeq20000{\rm km/s}$ shells (P. Mazzali, private comm.). Both Mazzali et al. and \citet{Tanaka09} find a low, $\sim0.01$, mass fraction of O at the fastest shells.

The comparison of our analysis of the early emission of SN2008D with the analyses of the light curve and spectra at maximum light indicates that a combined model, describing both the early emission from the expanding and cooling envelope and the emission at maximum light, which is driven by radioactive decay, will provide much better constraints on the progenitor and explosion parameters than those that may be obtained by analyzing either of the two separately.

\acknowledgements We thank P. Mazzali for useful discussions. This research was supported in part by ISF, AEC and Minerva grants.

\bibliographystyle{apj}
\bibliography{general}

\end{document}

%% file: Definitions.tex
\newcommand{\DDir}{\relax{D\kern-.7em{/}}}

\newcommand{\be}{\begin{equation}}
\newcommand{\ee}{\end{equation}}
\newcommand{\bea}{\begin{equation*}}
\newcommand{\eea}{\end{equation*}}

\newcommand{\nin}{\relax{\in\kern-.8em{/}}}

\newcommand{\gr}{\mbox{ g}}
\newcommand{\cm}{\mbox{ cm}}

\newcommand{\se}{\mbox{ s}}

\newcommand{\erg}{\mbox{ erg}}

\newcommand{\eV}{\mbox{ eV}}

\newcommand{\ph}{\mbox{ ph}}

\newcommand{\OpUnit}{\mbox{cm}^2 \mbox{g}^{-1} }
\newcommand{\DenUnit}{\mbox{g}\mbox{cm}^{-3}}
\newcommand{\TempUnit}{^{\circ}\mbox{K}}

\newcommand{\sref}{\S~\ref}

%% file: ms.bbl
\begin{thebibliography}{61}
\expandafter\ifx\csname natexlab\endcsname\relax\def\natexlab#1{#1}\fi

\bibitem[{{Band} {et~al.}(2008){Band}, {Grindlay}, {Hong}, {Fishman},
  {Hartmann}, {Garson}, {Krawczynski}, {Barthelmy}, {Gehrels}, \&
  {Skinner}}]{Band08}
{Band}, D.~L., {Grindlay}, J.~E., {Hong}, J., {Fishman}, G., {Hartmann}, D.~H.,
  {Garson}, III, A., {Krawczynski}, H., {Barthelmy}, S., {Gehrels}, N., \&
  {Skinner}, G. 2008, \apj, 673, 1225

\bibitem[{{Bernstein} \& {Book}(1980)}]{Bernstein80}
{Bernstein}, I.~B. \& {Book}, D.~L. 1980, \apj, 240, 223

\bibitem[{{Blinnikov} {et~al.}(2000){Blinnikov}, {Lundqvist}, {Bartunov},
  {Nomoto}, \& {Iwamoto}}]{Blinnikov00}
{Blinnikov}, S., {Lundqvist}, P., {Bartunov}, O., {Nomoto}, K., \& {Iwamoto},
  K. 2000, apj, 532, 1132

\bibitem[{{Blinnikov} {et~al.}(1998){Blinnikov}, {Eastman}, {Bartunov},
  {Popolitov}, \& {Woosley}}]{Blinnikov98}
{Blinnikov}, S.~I., {Eastman}, R., {Bartunov}, O.~S., {Popolitov}, V.~A., \&
  {Woosley}, S.~E. 1998, \apj, 496, 454

\bibitem[{{Blinnikov} {et~al.}(2002){Blinnikov}, {Nadyozhin}, {Woosley}, \&
  {Sorokina}}]{Blinnikov02}
{Blinnikov}, S.~I., {Nadyozhin}, D.~K., {Woosley}, S.~E., \& {Sorokina}, E.~I.
  2002, in Nuclear Astrophysics, ed. {W.~Hillebrandt \& E.~M{\"u}ller},
  144--147

\bibitem[{{Calzavara} \& {Matzner}(2004)}]{Calzavara04}
{Calzavara}, A.~J. \& {Matzner}, C.~D. 2004, \mnras, 351, 694

\bibitem[{{Campana} {et~al.}(2006){Campana}, {Mangano}, {Blustin}, {Brown},
  {Burrows}, {Chincarini}, {Cummings}, {Cusumano}, {Della Valle}, {Malesani},
  {M{\'e}sz{\'a}ros}, {Nousek}, {Page}, {Sakamoto}, {Waxman}, {Zhang}, {Dai},
  {Gehrels}, {Immler}, {Marshall}, {Mason}, {Moretti}, {O'Brien}, {Osborne},
  {Page}, {Romano}, {Roming}, {Tagliaferri}, {Cominsky}, {Giommi}, {Godet},
  {Kennea}, {Krimm}, {Angelini}, {Barthelmy}, {Boyd}, {Palmer}, {Wells}, \&
  {White}}]{EliNatur}
{Campana}, S., {Mangano}, V., {Blustin}, A.~J., {Brown}, P., {Burrows}, D.~N.,
  {Chincarini}, G., {Cummings}, J.~R., {Cusumano}, G., {Della Valle}, M.,
  {Malesani}, D., {M{\'e}sz{\'a}ros}, P., {Nousek}, J.~A., {Page}, M.,
  {Sakamoto}, T., {Waxman}, E., {Zhang}, B., {Dai}, Z.~G., {Gehrels}, N.,
  {Immler}, S., {Marshall}, F.~E., {Mason}, K.~O., {Moretti}, A., {O'Brien},
  P.~T., {Osborne}, J.~P., {Page}, K.~L., {Romano}, P., {Roming}, P.~W.~A.,
  {Tagliaferri}, G., {Cominsky}, L.~R., {Giommi}, P., {Godet}, O., {Kennea},
  J.~A., {Krimm}, H., {Angelini}, L., {Barthelmy}, S.~D., {Boyd}, P.~T.,
  {Palmer}, D.~M., {Wells}, A.~A., \& {White}, N.~E. 2006, Nature, 442, 1008

\bibitem[{{Cardelli} {et~al.}(1989){Cardelli}, {Clayton}, \&
  {Mathis}}]{cardelli}
{Cardelli}, J.~A., {Clayton}, G.~C., \& {Mathis}, J.~S. 1989, apj, 345, 245

\bibitem[{{Castor}(2004)}]{Castor04}
{Castor}, J.~I. 2004, {Radiation Hydrodynamics}, ed. {Castor, J.~I.}

\bibitem[{{Chandrasekhar}(1939)}]{Chandrasekhar39}
{Chandrasekhar}, S. 1939, {An introduction to the study of stellar structure},
  ed. S.~Chandrasekhar

\bibitem[{{Chevalier}(1992)}]{Chev}
{Chevalier}, R.~A. 1992, apj, 394, 599

\bibitem[{{Chevalier} \& {Fransson}(2008)}]{Chev08}
{Chevalier}, R.~A. \& {Fransson}, C. 2008, \apjl, 683, L135

\bibitem[{{Chevalier} \& {Soker}(1989)}]{Chevalier89}
{Chevalier}, R.~A. \& {Soker}, N. 1989, ApJ, 341, 867

\bibitem[{{Colgate}(1974)}]{Colgate74}
{Colgate}, S.~A. 1974, \apj, 187, 333

\bibitem[{{Crowther}(2007)}]{Crowther07}
{Crowther}, P.~A. 2007, \araa, 45, 177

\bibitem[{{Dessart} \& {Hillier}(2005)}]{Dessart05}
{Dessart}, L. \& {Hillier}, D.~J. 2005, \aap, 437, 667

\bibitem[{{Eastman} \& {Pinto}(1993)}]{Eastman93}
{Eastman}, R.~G. \& {Pinto}, P.~A. 1993, \apj, 412, 731

\bibitem[{{Ensman} \& {Burrows}(1992)}]{Ensman92}
{Ensman}, L. \& {Burrows}, A. 1992, \apj, 393, 742

\bibitem[{{Falk}(1978)}]{Falk78}
{Falk}, S.~W. 1978, \apjl, 225, L133

\bibitem[{{Fan} {et~al.}(2006){Fan}, {Piran}, \& {Xu}}]{Fan06}
{Fan}, Y., {Piran}, T., \& {Xu}, D. 2006, Journal of Cosmology and
  Astro-Particle Physics, 9, 13

\bibitem[{{Friend} \& {Castor}(1983)}]{Friend83}
{Friend}, D.~B. \& {Castor}, J.~I. 1983, \apj, 272, 259

\bibitem[{{Gaffet}(1984)}]{Gaffet84}
{Gaffet}, B. 1984, \aap, 135, 94

\bibitem[{{Gandel'Man} \& {Frank-Kamenetskii}(1956)}]{GandelMan56}
{Gandel'Man}, G.~M. \& {Frank-Kamenetskii}, D.~A. 1956, Soviet Physics Doklady,
  1, 223

\bibitem[{{Gezari} {et~al.}(2008){Gezari}, {Dessart}, {Basa}, {Martin},
  {Neill}, {Woosley}, {Hillier}, {Bazin}, {Forster}, {Friedman}, {Le Du},
  {Mazure}, {Morrissey}, {Neff}, {Schiminovich}, \& {Wyder}}]{Gezari08}
{Gezari}, S., {Dessart}, L., {Basa}, S., {Martin}, D.~C., {Neill}, J.~D.,
  {Woosley}, S.~E., {Hillier}, D.~J., {Bazin}, G., {Forster}, K., {Friedman},
  P.~G., {Le Du}, J., {Mazure}, A., {Morrissey}, P., {Neff}, S.~G.,
  {Schiminovich}, D., \& {Wyder}, T.~K. 2008, apjl, 683, L131

\bibitem[{{Ghisellini} {et~al.}(2007){Ghisellini}, {Ghirlanda}, \&
  {Tavecchio}}]{Ghisellini07}
{Ghisellini}, G., {Ghirlanda}, G., \& {Tavecchio}, F. 2007, \mnras, 382, L77

\bibitem[{{Grindlay} {et~al.}(2003){Grindlay}, {Craig}, {Gehrels}, {Harrison},
  \& {Hong}}]{Grindlay03}
{Grindlay}, J.~E., {Craig}, W.~W., {Gehrels}, N.~A., {Harrison}, F.~A., \&
  {Hong}, J. 2003, in Society of Photo-Optical Instrumentation Engineers (SPIE)
  Conference Series, Vol. 4851, Society of Photo-Optical Instrumentation
  Engineers (SPIE) Conference Series, ed. {J.~E.~Truemper \& H.~D.~Tananbaum},
  331--344

\bibitem[{{Hauschildt} \& {Ensman}(1994)}]{Hauschildt94}
{Hauschildt}, P.~H. \& {Ensman}, L.~M. 1994, \apj, 424, 905

\bibitem[{{Karp} {et~al.}(1977){Karp}, {Lasher}, {Chan}, \&
  {Salpeter}}]{karp77}
{Karp}, A.~H., {Lasher}, G., {Chan}, K.~L., \& {Salpeter}, E.~E. 1977, \apj,
  214, 161

\bibitem[{{Katz} {et~al.}(2010){Katz}, {Budnik}, \& {Waxman}}]{Katz10}
{Katz}, B., {Budnik}, R., \& {Waxman}, E. 2010, \apj, 716, 781

\bibitem[{{Klein} \& {Chevalier}(1978)}]{Klein78}
{Klein}, R.~I. \& {Chevalier}, R.~A. 1978, \apjl, 223, L109

\bibitem[{{Law} {et~al.}(2009){Law}, {Kulkarni}, {Dekany}, {Ofek}, {Quimby},
  {Nugent}, {Surace}, {Grillmair}, {Bloom}, {Kasliwal}, {Bildsten}, {Brown},
  {Cenko}, {Ciardi}, {Croner}, {Djorgovski}, {van Eyken}, {Filippenko}, {Fox},
  {Gal-Yam}, {Hale}, {Hamam}, {Helou}, {Henning}, {Howell}, {Jacobsen},
  {Laher}, {Mattingly}, {McKenna}, {Pickles}, {Poznanski}, {Rahmer}, {Rau},
  {Rosing}, {Shara}, {Smith}, {Starr}, {Sullivan}, {Velur}, {Walters}, \&
  {Zolkower}}]{Law09}
{Law}, N.~M., {Kulkarni}, S.~R., {Dekany}, R.~G., {Ofek}, E.~O., {Quimby},
  R.~M., {Nugent}, P.~E., {Surace}, J., {Grillmair}, C.~C., {Bloom}, J.~S.,
  {Kasliwal}, M.~M., {Bildsten}, L., {Brown}, T., {Cenko}, S.~B., {Ciardi}, D.,
  {Croner}, E., {Djorgovski}, S.~G., {van Eyken}, J., {Filippenko}, A.~V.,
  {Fox}, D.~B., {Gal-Yam}, A., {Hale}, D., {Hamam}, N., {Helou}, G., {Henning},
  J., {Howell}, D.~A., {Jacobsen}, J., {Laher}, R., {Mattingly}, S., {McKenna},
  D., {Pickles}, A., {Poznanski}, D., {Rahmer}, G., {Rau}, A., {Rosing}, W.,
  {Shara}, M., {Smith}, R., {Starr}, D., {Sullivan}, M., {Velur}, V.,
  {Walters}, R., \& {Zolkower}, J. 2009, \pasp, 121, 1395

\bibitem[{{Li}(2007)}]{Li07}
{Li}, L. 2007, \mnras, 375, 240

\bibitem[{{Li}(2008)}]{Li08}
---. 2008, \mnras, 388, 603

\bibitem[{{Maeda} {et~al.}(2007){Maeda}, {Kawabata}, {Tanaka}, {Nomoto},
  {Tominaga}, {Hattori}, {Minezaki}, {Kuroda}, {Suzuki}, {Deng}, {Mazzali}, \&
  {Pian}}]{Maeda07Apj}
{Maeda}, K., {Kawabata}, K., {Tanaka}, M., {Nomoto}, K., {Tominaga}, N.,
  {Hattori}, T., {Minezaki}, T., {Kuroda}, T., {Suzuki}, T., {Deng}, J.,
  {Mazzali}, P.~A., \& {Pian}, E. 2007, apjl, 658, L5

\bibitem[{{Malesani} {et~al.}(2009){Malesani}, {Fynbo}, {Hjorth}, {Leloudas},
  {Sollerman}, {Stritzinger}, {Vreeswijk}, {Watson}, {Gorosabel},
  {Micha{\l}owski}, {Th{\"o}ne}, {Augusteijn}, {Bersier}, {Jakobsson},
  {Jaunsen}, {Ledoux}, {Levan}, {Milvang-Jensen}, {Rol}, {Tanvir}, {Wiersema},
  {Xu}, {Albert}, {Bayliss}, {Gall}, {Grove}, {Koester}, {Leitet}, {Pursimo},
  \& {Skillen}}]{Malesani09}
{Malesani}, D., {Fynbo}, J.~P.~U., {Hjorth}, J., {Leloudas}, G., {Sollerman},
  J., {Stritzinger}, M.~D., {Vreeswijk}, P.~M., {Watson}, D.~J., {Gorosabel},
  J., {Micha{\l}owski}, M.~J., {Th{\"o}ne}, C.~C., {Augusteijn}, T., {Bersier},
  D., {Jakobsson}, P., {Jaunsen}, A.~O., {Ledoux}, C., {Levan}, A.~J.,
  {Milvang-Jensen}, B., {Rol}, E., {Tanvir}, N.~R., {Wiersema}, K., {Xu}, D.,
  {Albert}, L., {Bayliss}, M., {Gall}, C., {Grove}, L.~F., {Koester}, B.~P.,
  {Leitet}, E., {Pursimo}, T., \& {Skillen}, I. 2009, ApJ, 692, L84

\bibitem[{{Matsuoka} {et~al.}(1997){Matsuoka}, {Kawai}, {Mihara}, {Yoshida},
  {Kubo}, {Kotani}, {Negoro}, {Rubin}, {Shimizu}, {Tsunemi}, {Hayashida},
  {Kitamoto}, {Miyata}, \& {Yamauchi}}]{Matsuoka97}
{Matsuoka}, M., {Kawai}, N., {Mihara}, T., {Yoshida}, A., {Kubo}, H., {Kotani},
  T., {Negoro}, H., {Rubin}, B.~C., {Shimizu}, H.~M., {Tsunemi}, H.,
  {Hayashida}, K., {Kitamoto}, S., {Miyata}, E., \& {Yamauchi}, M. 1997, in
  Society of Photo-Optical Instrumentation Engineers (SPIE) Conference Series,
  Vol. 3114, Society of Photo-Optical Instrumentation Engineers (SPIE)
  Conference Series, ed. {O.~H.~Siegmund \& M.~A.~Gummin}, 414--421

\bibitem[{{Matzner} \& {McKee}(1999)}]{MM}
{Matzner}, C.~D. \& {McKee}, C.~F. 1999, apj, 510, 379

\bibitem[{{Mazzali} {et~al.}(2006){Mazzali}, {Deng}, {Nomoto}, {Sauer}, {Pian},
  {Tominaga}, {Tanaka}, {Maeda}, \& {Filippenko}}]{Mazzali06Nat}
{Mazzali}, P.~A., {Deng}, J., {Nomoto}, K., {Sauer}, D.~N., {Pian}, E.,
  {Tominaga}, N., {Tanaka}, M., {Maeda}, K., \& {Filippenko}, A.~V. 2006,
  Nature, 442, 1018

\bibitem[{{Mazzali} {et~al.}(2008){Mazzali}, {Valenti}, {Della Valle},
  {Chincarini}, {Sauer}, {Benetti}, {Pian}, {Piran}, {D'Elia}, {Elias-Rosa},
  {Margutti}, {Pasotti}, {Antonelli}, {Bufano}, {Campana}, {Cappellaro},
  {Covino}, {D'Avanzo}, {Fiore}, {Fugazza}, {Gilmozzi}, {Hunter}, {Maguire},
  {Maiorano}, {Marziani}, {Masetti}, {Mirabel}, {Navasardyan}, {Nomoto},
  {Palazzi}, {Pastorello}, {Panagia}, {Pellizza}, {Sari}, {Smartt},
  {Tagliaferri}, {Tanaka}, {Taubenberger}, {Tominaga}, {Trundle}, \&
  {Turatto}}]{Mazzali08}
{Mazzali}, P.~A., {Valenti}, S., {Della Valle}, M., {Chincarini}, G., {Sauer},
  D.~N., {Benetti}, S., {Pian}, E., {Piran}, T., {D'Elia}, V., {Elias-Rosa},
  N., {Margutti}, R., {Pasotti}, F., {Antonelli}, L.~A., {Bufano}, F.,
  {Campana}, S., {Cappellaro}, E., {Covino}, S., {D'Avanzo}, P., {Fiore}, F.,
  {Fugazza}, D., {Gilmozzi}, R., {Hunter}, D., {Maguire}, K., {Maiorano}, E.,
  {Marziani}, P., {Masetti}, N., {Mirabel}, F., {Navasardyan}, H., {Nomoto},
  K., {Palazzi}, E., {Pastorello}, A., {Panagia}, N., {Pellizza}, L.~J.,
  {Sari}, R., {Smartt}, S., {Tagliaferri}, G., {Tanaka}, M., {Taubenberger},
  S., {Tominaga}, N., {Trundle}, C., \& {Turatto}, M. 2008, Science, 321, 1185

\bibitem[{{Meynet} \& {Maeder}(2003)}]{Meynet03}
{Meynet}, G. \& {Maeder}, A. 2003, aap, 404, 975

\bibitem[{{Mihalas} \& {Mihalas}(1984)}]{Mihalas84}
{Mihalas}, D. \& {Mihalas}, B.~W. 1984, {Foundations of radiation
  hydrodynamics}, ed. B.~W. Mihalas, D. \&~Mihalas

\bibitem[{{Modjaz} {et~al.}(2009){Modjaz}, {Li}, {Butler}, {Chornock},
  {Perley}, {Blondin}, {Bloom}, {Filippenko}, {Kirshner}, {Kocevski},
  {Poznanski}, {Hicken}, {Foley}, {Stringfellow}, {Berlind}, {Barrado y
  Navascues}, {Blake}, {Bouy}, {Brown}, {Challis}, {Chen}, {de Vries},
  {Dufour}, {Falco}, {Friedman}, {Ganeshalingam}, {Garnavich}, {Holden},
  {Illingworth}, {Lee}, {Liebert}, {Marion}, {Olivier}, {Prochaska},
  {Silverman}, {Smith}, {Starr}, {Steele}, {Stockton}, {Williams}, \&
  {Wood-Vasey}}]{Modjaz09}
{Modjaz}, M., {Li}, W., {Butler}, N., {Chornock}, R., {Perley}, D., {Blondin},
  S., {Bloom}, J.~S., {Filippenko}, A.~V., {Kirshner}, R.~P., {Kocevski}, D.,
  {Poznanski}, D., {Hicken}, M., {Foley}, R.~J., {Stringfellow}, G.~S.,
  {Berlind}, P., {Barrado y Navascues}, D., {Blake}, C.~H., {Bouy}, H.,
  {Brown}, W.~R., {Challis}, P., {Chen}, H., {de Vries}, W.~H., {Dufour}, P.,
  {Falco}, E., {Friedman}, A., {Ganeshalingam}, M., {Garnavich}, P., {Holden},
  B., {Illingworth}, G., {Lee}, N., {Liebert}, J., {Marion}, G.~H., {Olivier},
  S.~S., {Prochaska}, J.~X., {Silverman}, J.~M., {Smith}, N., {Starr}, D.,
  {Steele}, T.~N., {Stockton}, A., {Williams}, G.~G., \& {Wood-Vasey}, W.~M.
  2009, \apj, 702, 226

\bibitem[{{Modjaz} {et~al.}(2006){Modjaz}, {Stanek}, {Garnavich}, {Berlind},
  {Blondin}, {Brown}, {Calkins}, {Challis}, {Diamond-Stanic}, {Hao}, {Hicken},
  {Kirshner}, \& {Prieto}}]{Modjaz07}
{Modjaz}, M., {Stanek}, K.~Z., {Garnavich}, P.~M., {Berlind}, P., {Blondin},
  S., {Brown}, W., {Calkins}, M., {Challis}, P., {Diamond-Stanic}, A.~M.,
  {Hao}, H., {Hicken}, M., {Kirshner}, R.~P., \& {Prieto}, J.~L. 2006, ApJ,
  645, L21

\bibitem[{{Nomoto} {et~al.}(1995){Nomoto}, {Iwamoto}, \& {Suzuki}}]{Nomoto95}
{Nomoto}, K.~I., {Iwamoto}, K., \& {Suzuki}, T. 1995, \physrep, 256, 173

\bibitem[{{Pian} {et~al.}(2006){Pian}, {Mazzali}, {Masetti}, {Ferrero},
  {Klose}, {Palazzi}, {Ramirez-Ruiz}, {Woosley}, {Kouveliotou}, {Deng},
  {Filippenko}, {Foley}, {Fynbo}, {Kann}, {Li}, {Hjorth}, {Nomoto}, {Patat},
  {Sauer}, {Sollerman}, {Vreeswijk}, {Guenther}, {Levan}, {O'Brien}, {Tanvir},
  {Wijers}, {Dumas}, {Hainaut}, {Wong}, {Baade}, {Wang}, {Amati}, {Cappellaro},
  {Castro-Tirado}, {Ellison}, {Frontera}, {Fruchter}, {Greiner}, {Kawabata},
  {Ledoux}, {Maeda}, {M{\o}ller}, {Nicastro}, {Rol}, \& {Starling}}]{Pian06Nat}
{Pian}, E., {Mazzali}, P.~A., {Masetti}, N., {Ferrero}, P., {Klose}, S.,
  {Palazzi}, E., {Ramirez-Ruiz}, E., {Woosley}, S.~E., {Kouveliotou}, C.,
  {Deng}, J., {Filippenko}, A.~V., {Foley}, R.~J., {Fynbo}, J.~P.~U., {Kann},
  D.~A., {Li}, W., {Hjorth}, J., {Nomoto}, K., {Patat}, F., {Sauer}, D.~N.,
  {Sollerman}, J., {Vreeswijk}, P.~M., {Guenther}, E.~W., {Levan}, A.,
  {O'Brien}, P., {Tanvir}, N.~R., {Wijers}, R.~A.~M.~J., {Dumas}, C.,
  {Hainaut}, O., {Wong}, D.~S., {Baade}, D., {Wang}, L., {Amati}, L.,
  {Cappellaro}, E., {Castro-Tirado}, A.~J., {Ellison}, S., {Frontera}, F.,
  {Fruchter}, A.~S., {Greiner}, J., {Kawabata}, K., {Ledoux}, C., {Maeda}, K.,
  {M{\o}ller}, P., {Nicastro}, L., {Rol}, E., \& {Starling}, R. 2006, Nature,
  442, 1011

\bibitem[{{Quimby}(2006)}]{quimby_phd}
{Quimby}, R.~M. 2006, PhD thesis, University of Texas, United States -- Texas

\bibitem[{{Rabinak} {et~al.}(2010){Rabinak}, {Kushnir}, \&
  {Waxman}}]{RabinakIP}
{Rabinak}, I., {Kushnir}, D., \& {Waxman}, E. 2010, In preperation

\bibitem[{Sakurai(1960)}]{Sakurai60}
Sakurai, A. 1960, Communications on Pure and Applied Mathematics, 13, 353

\bibitem[{{Schawinski} {et~al.}(2008){Schawinski}, {Justham}, {Wolf},
  {Podsiadlowski}, {Sullivan}, {Steenbrugge}, {Bell}, {R{\"o}ser}, {Walker},
  {Astier}, {Balam}, {Balland}, {Carlberg}, {Conley}, {Fouchez}, {Guy},
  {Hardin}, {Hook}, {Howell}, {Pain}, {Perrett}, {Pritchet}, {Regnault}, \&
  {Yi}}]{Schawinski08}
{Schawinski}, K., {Justham}, S., {Wolf}, C., {Podsiadlowski}, P., {Sullivan},
  M., {Steenbrugge}, K.~C., {Bell}, T., {R{\"o}ser}, H.-J., {Walker}, E.~S.,
  {Astier}, P., {Balam}, D., {Balland}, C., {Carlberg}, R., {Conley}, A.,
  {Fouchez}, D., {Guy}, J., {Hardin}, D., {Hook}, I., {Howell}, D.~A., {Pain},
  R., {Perrett}, K., {Pritchet}, C., {Regnault}, N., \& {Yi}, S.~K. 2008,
  Science, 321, 223

\bibitem[{{Seaton}(2005)}]{OPCD}
{Seaton}, M.~J. 2005, mnras, 362, L1

\bibitem[{{Sedov}(1959)}]{Sedov59}
{Sedov}, L.~I. 1959, {Similarity and Dimensional Methods in Mechanics}, ed.
  L.~I. Sedov

\bibitem[{{Soderberg} {et~al.}(2008){Soderberg}, {Berger}, {Page}, {Schady},
  {Parrent}, {Pooley}, {Wang}, {Ofek}, {Cucchiara}, {Rau}, {Waxman}, {Simon},
  {Bock}, {Milne}, {Page}, {Barentine}, {Barthelmy}, {Beardmore}, {Bietenholz},
  {Brown}, {Burrows}, {Burrows}, {Byrngelson}, {Cenko}, {Chandra}, {Cummings},
  {Fox}, {Gal-Yam}, {Gehrels}, {Immler}, {Kasliwal}, {Kong}, {Krimm},
  {Kulkarni}, {Maccarone}, {M{\'e}sz{\'a}ros}, {Nakar}, {O'Brien}, {Overzier},
  {de Pasquale}, {Racusin}, {Rea}, \& {York}}]{Alicia08D}
{Soderberg}, A.~M., {Berger}, E., {Page}, K.~L., {Schady}, P., {Parrent}, J.,
  {Pooley}, D., {Wang}, X., {Ofek}, E.~O., {Cucchiara}, A., {Rau}, A.,
  {Waxman}, E., {Simon}, J.~D., {Bock}, D., {Milne}, P.~A., {Page}, M.~J.,
  {Barentine}, J.~C., {Barthelmy}, S.~D., {Beardmore}, A.~P., {Bietenholz},
  M.~F., {Brown}, P., {Burrows}, A., {Burrows}, D.~N., {Byrngelson}, G.,
  {Cenko}, S.~B., {Chandra}, P., {Cummings}, J.~R., {Fox}, D.~B., {Gal-Yam},
  A., {Gehrels}, N., {Immler}, S., {Kasliwal}, M., {Kong}, A.~K.~H., {Krimm},
  H.~A., {Kulkarni}, S.~R., {Maccarone}, T.~J., {M{\'e}sz{\'a}ros}, P.,
  {Nakar}, E., {O'Brien}, P.~T., {Overzier}, R.~A., {de Pasquale}, M.,
  {Racusin}, J., {Rea}, N., \& {York}, D.~G. 2008, \nat, 453, 469

\bibitem[{{Soderberg} {et~al.}(2006){Soderberg}, {Kulkarni}, {Nakar}, {Berger},
  {Cameron}, {Fox}, {Frail}, {Gal-Yam}, {Sari}, {Cenko}, {Kasliwal},
  {Chevalier}, {Piran}, {Price}, {Schmidt}, {Pooley}, {Moon}, {Penprase},
  {Ofek}, {Rau}, {Gehrels}, {Nousek}, {Burrows}, {Persson}, \&
  {McCarthy}}]{Soderberg06}
{Soderberg}, A.~M., {Kulkarni}, S.~R., {Nakar}, E., {Berger}, E., {Cameron},
  P.~B., {Fox}, D.~B., {Frail}, D., {Gal-Yam}, A., {Sari}, R., {Cenko}, S.~B.,
  {Kasliwal}, M., {Chevalier}, R.~A., {Piran}, T., {Price}, P.~A., {Schmidt},
  B.~P., {Pooley}, G., {Moon}, D., {Penprase}, B.~E., {Ofek}, E., {Rau}, A.,
  {Gehrels}, N., {Nousek}, J.~A., {Burrows}, D.~N., {Persson}, S.~E., \&
  {McCarthy}, P.~J. 2006, \nat, 442, 1014

\bibitem[{{Tanaka} {et~al.}(2009){Tanaka}, {Tominaga}, {Nomoto}, {Valenti},
  {Sahu}, {Minezaki}, {Yoshii}, {Yoshida}, {Anupama}, {Benetti}, {Chincarini},
  {Valle}, {Mazzali}, \& {Pian}}]{Tanaka09}
{Tanaka}, M., {Tominaga}, N., {Nomoto}, K., {Valenti}, S., {Sahu}, D.~K.,
  {Minezaki}, T., {Yoshii}, Y., {Yoshida}, M., {Anupama}, G.~C., {Benetti}, S.,
  {Chincarini}, G., {Valle}, M.~D., {Mazzali}, P.~A., \& {Pian}, E. 2009, ApJ,
  692, 1131

\bibitem[{{Taylor}(1950)}]{Taylor50}
{Taylor}, G. 1950, Royal Society of London Proceedings Series A, 201, 175

\bibitem[{{Von Neumann}(1947)}]{VonNeumann47}
{Von Neumann}, J. 1947, {Blast Waves}, Los alamos sci. lab. tech. series, vol.
  7, Los Alamos, NM

\bibitem[{{Wagoner} {et~al.}(1991){Wagoner}, {Perez}, \& {Vasu}}]{Wagoner91}
{Wagoner}, R.~V., {Perez}, C.~A., \& {Vasu}, M. 1991, \apj, 377, 639

\bibitem[{{Waxman} {et~al.}(2007){Waxman}, {M{\'e}sz{\'a}ros}, \&
  {Campana}}]{Waxman07}
{Waxman}, E., {M{\'e}sz{\'a}ros}, P., \& {Campana}, S. 2007, \apj, 667, 351

\bibitem[{{Weaver} {et~al.}(1978){Weaver}, {Zimmerman}, \&
  {Woosley}}]{Weaver78}
{Weaver}, T.~A., {Zimmerman}, G.~B., \& {Woosley}, S.~E. 1978, \apj, 225, 1021

\bibitem[{{Woosley} {et~al.}(1993){Woosley}, {Langer}, \& {Weaver}}]{Woosley93}
{Woosley}, S.~E., {Langer}, N., \& {Weaver}, T.~A. 1993, \apj, 411, 823

\bibitem[{{Woosley} \& {Weaver}(1986)}]{WW86}
{Woosley}, S.~E. \& {Weaver}, T.~A. 1986, \araa, 24, 205

\end{thebibliography}
